\documentclass[12pt]{article}
\usepackage[nonatbib,preprint]{neurips_2020}
\usepackage[numbers,compress]{natbib}
\usepackage{booktabs}       % professional-quality tables
\usepackage{amsfonts}       % blackboard math symbols
\usepackage{nicefrac}       % compact symbols for 1/2, etc.
\usepackage{graphicx}
\usepackage[dvipsnames]{xcolor}
\usepackage{enumitem}
\usepackage{rotating}
\usepackage{array}
\usepackage{mathtools}

\usepackage{wrapfig}
\usepackage{url,lineno,microtype,subcaption}
\usepackage[onehalfspacing]{setspace}
\usepackage{blindtext}
\usepackage[ruled,vlined]{algorithm2e}
\usepackage{cite}
\usepackage{bm}
\newcommand{\ind}{\perp\!\!\!\!\perp}
\usepackage{amsmath,amsthm}
\theoremstyle{definition}
\newtheorem{exmp}{Example}[section]
\newtheorem{corthm}{Corollary}[section]
\newtheorem{theorem}{Theorem}

\makeatletter
 \def\@textbottom{\vskip \z@ \@plus 1pt}
 \let\@texttop\relax
\makeatother
\newtheorem{definition}{Definition}
\usepackage[final]{hyperref} % adds hyper links inside the generated pdf file
\hypersetup{
	colorlinks=true,       % false: boxed links; true: colored links
	linkcolor=blue,        % color of internal links
	citecolor=blue,        % color of links to bibliography
	filecolor=magenta,     % color of file links
	urlcolor=blue         
}

\title{Statistical Perspective on Functional and Causal Neural Connectomics: The Time-Aware PC Algorithm}
\author{
  Rahul Biswas\\ 
  Department of Statistics\\
  University of Washington\\
  Seattle, WA, 98195\\
  \texttt{rbiswas1@uw.edu}\\
  \And
  Eli Shlizerman \\ 
  Department of Applied Mathematics\\
  Department of Electrical \& Computer Engineering\\  University of Washington\\
  Seattle, WA, 98195 \\
  \texttt{shlizee@uw.edu} \\
}

\begin{document}
\maketitle
\begin{abstract}
The representation of the flow of information between neurons in the brain based on their activity is termed the \emph{causal functional connectome}. Such representation incorporates the dynamic nature of neuronal activity and causal interactions between them. In contrast to connectome, the causal functional connectome is not directly observed and needs to be inferred from neural time series. A popular statistical framework for inferring causal connectivity from observations is the \emph{directed probabilistic graphical modeling}. Its common formulation is not suitable for neural time series since was developed for variables with independent and identically distributed static samples. In this work, we propose to model and estimate the causal functional connectivity from neural time series using a novel approach that adapts directed probabilistic graphical modeling to the time series scenario. In particular, we develop the \emph{Time-Aware PC} (TPC) algorithm for estimating the causal functional connectivity, which adapts the PC algorithm a state-of-the-art method for statistical causal inference. We show that the model outcome of TPC has the properties of reflecting causality of neural interactions such as being non-parametric, exhibits the \emph{directed Markov} property in a time-series setting, and is predictive of the consequence of counterfactual interventions on the time series. We demonstrate the utility of the methodology to obtain the causal functional connectome for several datasets including simulations, benchmark datasets, and recent multi-array electro-physiological recordings from the mouse visual cortex.
\end{abstract}

\section{Introduction}
\emph{Functional Connectome} (FC) refers to the network of interactions between units of the brain, such as individual neurons or brain regions, with respect to their activity over time \citep{reid2012functional}. The aim of finding the FC is to provide insight into how neurons interact to form brain function. FC can be represented by a graph whose nodes represent neurons and edges indicate a relationship between the activity of connected neurons. The edges can either represent undirected stochastic associations between activity of neurons or directed causal relationships between activity of neurons. While association between neural activity describes whether neuron $A$ and neuron $B$ are active in a correlated manner, the ultimate goal of functional connectomics is to answer causal queries, such as whether the activity in neuron $A$ causes neuron $B$ to be active $(A\rightarrow B)$, or is it the other way around $(B\rightarrow A)$? Else, does a neuron $C$ intermediate the correlation between $A$ and $B$ $(A\leftarrow C \rightarrow B)$ \citep{reid2019advancing,cassidy2021functional,sanchez2021combining}?

When the interactions are causal, the network is termed as \emph{causal functional connectome} (CFC). The CFC maps how neural activity flows within neural circuits, and provides the possibility for inference of neural pathways essential for brain functioning and behavior, such as sensory-motor-behavioral pathways \citep{finn2015functional}. Several approaches aim to infer CFC, such as Granger Causality (GC), Dynamic Causal Modeling (DCM), and Directed Probabilistic Graphical Models (DPGM), each having their applicability and challenges, as surveyed in \citep{biswas2021statistical}. GC obtains the directed functional connectivity from observed neural activity in a way that tells whether a neuron’s \emph{past} is predictive of another neuron's future, however it is unclear whether the prediction implies causation. In contrast, DCM compares \emph{specific} mechanistic biological models based on data evidence, in which, model parameters represent causal influences between hidden neural states \citep{sharaev2016effective}. On the other hand, DPGM is a generic procedure to obtain causal relationships between nodes of a network from observations, using the \emph{directed Markov} property. DPGM is non-parametric in the sense of capturing arbitrary functional relationships among the nodes and is predictive of the consequence of counterfactual interventions to the network \citep{lauritzen2001causal,maathuis2018handbook}. Such properties make DPGM a popular approach for causal modeling in various disciplines such as genomics and econometrics~\citep{gomez2020functional,  ahelegbey2016econometrics, ebert2012causal, kalisch2010understanding, deng2005structural, haigh2004causality,wang2017potential,sinoquet2014probabilistic, mourad2012probabilistic, wang2005new, liu2018functional, friedman2004inferring}. 

The utility of DPGM in obtaining CFC from neural data has been investigated in \citep{biswas2021statistical}. DPGM, applicable to i.i.d. observations, can model causal relations between whole time series of different neurons in sense of average over time or at a specific time. Thereby, standard DPGM does not explicitly model the inter-temporal causal relations between neural activity in the time series. For inference of the DPGM, the PC algorithm is one of the widely used causal inference algorithms that assumes independent and identically distributed (i.i.d.) sampling of the nodes of the network and absence of latent confounders \citep{spirtes2000causation,pearl2009causality}. However, in neural time series scenario, causal relations are between neural activity at different times. Assuming independent sampling of nodes of the common DPGM is not suitable as the nodes correspond to a time series with temporal dependency. Moreover, DPGM typically generate a Directed Acyclic Graph (DAG), while neural activity is often comprised of feedback loops over time \citep{fiete2010spike,arbabshirani2019autoconnectivity,borisyuk1999oscillatory,jutras2010synchronous}. Though adaptations aim to include cycles in the DPGM they have a more complicated output  \citep{richardson1996automated}. Addressing these limitations will improve the utility of DPGM for finding CFC in the neural time series scenario and is the focus of this work.

In this work, we develop a novel approach for modeling and estimating causal functional connectivity by adapting directed probabilistic graphical models to the time series scenario. We introduce the \emph{Time-Aware PC} (TPC) algorithm. It uses the PC algorithm as a starting point and adapts it to the neural time series setup by following processes such as time-delay, bootstrapping and pruning. These ensure that the inferred CFC is robust and well suited to the setup. The proposed CFC graphical model incorporates feedback loops in functional connectivity and is non-parametric, yet we show that the CFC graphical model accurately represents the causal relationships in the unknown dynamical process of neural activity. Furthermore, the proposed CFC graphical model is predictive of the consequence of counterfactual interventions, such as the alteration in the CFC due to ablation or external control of certain neurons. We apply the proposed methodology on neural signals simulated from different paradigms and CFC motifs and demonstrate the utility of TPC in recovering the generating motifs. We also apply the TPC on public benchmark datasets and compare the performance in recovery of the ground truth CFC with other approaches. We further demonstrate the use of TPC to obtain the CFC among sampled neurons in mice brain from electrophysiological neural signals.

The following is a list of acronyms used in this paper: Functional Connectivity (FC), Causal Functional Connectivity (CFC), Granger Causality (GC), Dynamic Causal Model (DCM), Directed Probabilistic Graphical Model (DPGM), Directed Markov Property (DMP), Functional Magnetic Resonance Imaging (fMRI), Probabilistic Graphical Model (PGM), i.i.d. (Independent and Identically Distributed), Markov Property (MP), Directed Acyclic Graph (DAG), Peter Clark (PC), Greedy Equivalence Search (GES), Greedy Interventional Equivalence Search (GIES), Continuous Time Recurrent Neural Network (CTRNN), True Positive (TP), False Positive (FP), True Negative (TN), False Negative (FN), True Positive Rate (TPR), False Positive Rate (FPR).

\section{Causal Functional Connectivity for Static Variables - Review}\label{sec:dgm}

In this section, we provide a concise summary of DPGM for finding CFC for static variables, i.e. variables with \textbf{i.i.d. samples}, and extend in later sections to incorporate temporal dependence in time series. Let us consider a brain network $V=\{v_1,\ldots, v_N\}$ with $N$ neurons labeled as $v_1,\ldots,v_N$ and $X_v (t) \in \mathbb{R}$ denote a random variable measuring the activity of neuron $v$ at time $t$. Examples for such variables are instantaneous membrane potential, instantaneous firing rate, etc. Let $Y_v$ denote a scalar-valued random variable corresponding to $v\in V$, e.g., the neural recording at time $t$: $Y_v=X_v(t)$, average of recordings over time $Y_v=\bar{X}(v)$, and for a set of neurons $A\subset V$, $\bm{Y}_A$ denotes the random vector $(Y_v, v\in A)$. Let $G=(V,E)$ denote a \emph{directed acyclic graph} (DAG), i.e., a directed graph without directed cycles, over the neurons in $V$ and with directed edges $E$. Nodes $u$ and $v \in V$ are said to be \emph{adjacent} if $v\rightarrow u \in E$ or $u\rightarrow v \in E$. A \emph{path} is a sequence of distinct nodes in which successive nodes are adjacent. For a path $\pi = (v_0,\ldots, v_k)$, if every edge of $\pi$ is of the form $v_{i-1}\rightarrow v_{i}$ then $v_0$ is an \emph{ancestor} of $v_k$ and $v_k$ is a \emph{descendant} of $v_0$. The set of \emph{non-descendants} of $v$, denoted $nd_G(v)$, contains nodes $u\in V\setminus \{v\}$ that are not descendants of $v$. The set of \emph{parents} of $v\in V$ is denoted as $pa_G(v) =\{u\in V: u\rightarrow v\in E\}$. We mark the set $nd_G (v)\setminus pa_G (v)$ as the set that contains all nodes which are older ancestors of $v$ before its parents  \citep{lauritzen2001causal,maathuis2018handbook}. We use the convention that $u-v \in G \Rightarrow u\rightarrow v \in G$ and $u\leftarrow v \in G$. 

With these notations, we highlight the \emph{Directed Markov Property} (DMP) and its functional equivalence for DPGM. The DMP connects probabilistic conditional independencies between nodes of a directed graph with relationships of causal influence by ensuring that the influence of each node's ancestors beyond parents reaches to the node exclusively via its parents. Furthermore, the functional equivalence for DPGM shows that, the edges in a DPGM satisfying the DMP are consistent with causal functional interactions among the nodes.

\paragraph{Directed Markov Property (DMP)}
$(Y_v,v\in V)$ is said to satisfy the \emph{Directed Markov Property} with respect to the DAG $G$ if and only if,
\begin{equation}\label{eq:dgmarkov}
Y_v\ind \bm{Y}_{nd_G (v)\setminus pa_G (v)}|\bm{Y}_{pa_G (v)}
\end{equation}

The DMP translates the edges in the DAG into conditional independencies, such that each node $Y_v$ and its older ancestors $\bm{Y}_{nd_G (v)\setminus pa_G (v)}$ are conditionally independent given its parents $\bm{Y}_{pa_G}(v)$. The DMP can be equivalently represented with functional relationships between parent and child instead of conditional independencies, which is described in the following theorem \citep{bollen1989structural}. 

\paragraph{Functional Equivalence of DMP}\label{thm:fnequiv}
If $Y_v$ satisfies
\begin{equation}\label{eq:causaleff-structural}Y_v = g_v(Y_{pa_{\tilde{G}}(v)}, \epsilon_v), v\in V\end{equation}
where $\epsilon_v$ are independent random variables and $g_v$ are measurable functions for $v\in V$ and $\tilde{G}$ is a DAG with vertices $V$, then $Y_v, v\in V$ satisfies the Directed Markov Property with respect to $\tilde{G}$. Conversely, if $Y_v,v\in V$ satisfies the Directed Markov Property with respect to a DAG $\tilde{G}$, then there are
independent random variables $\epsilon_v$ and measurable functions $g_v$ for which Eq. (\ref{eq:causaleff-structural}) holds. This shows that if $Y_v, v\in V$ satisfies the DMP with respect to the DAG $G$, then $G$ admits a natural causal interpretation, due to its functional equivalence: parent nodes of $v$ in $G$ causally influence the child node $v$ \citep{drton2017structure}.

\paragraph{PC algorithm} Let $Y_v, v\in V$ satisfy the DMP with respect to the DAG $G$. The PC algorithm is a popular method to infer $G$ from observed data \citep{spirtes2000causation}. The PC algorithm uses a consistent statistical test, such as Fisher's Z-transform when $Y_v, v\in V$ are Gaussian variables, and kernel and distance based tests for non-Gaussian variables \citep{kalisch2007estimating,tillman2009nonlinear}. The algorithm first represents the observed variables by nodes of a graph and starts with an empty set of edges and puts an undirected edge between each pair of nodes if they are independent or conditionally independent given any other variable(s) determined by the statistical test. This results in the undirected skeleton graph, which is then converted into a DAG by directing the undirected edges using rules for orientation. The PC algorithm estimates several DAGs $\hat{G}_{i}$ based on i.i.d. samples of $Y_v,v\in V$, and outputs a single completed partially directed acyclic graph (CPDAG) $\hat{G}$ defined as follows: $\hat{G}$ has a directed edge from node $v \rightarrow w$ if $v\rightarrow w$ is present in all the DAGs $\hat{G}_{i}$. It has an undirected edge between $v$ and $w$ if either directions between them are present among the DAGs $\hat{G}_{i}$. It has no edge between $v$ and $w$ if no edge is present between them in any of the DAGs $\hat{G}_{i}$. The CPDAG $\hat{G}$ is uniquely identifiable from observed data. 

The PC algorithm assumes \emph{causal sufficiency} of the input variables: A set $V$ of variables is causally sufficient for a population if and only if in the population every common cause of any two or more variables in $V$ is in $V$, or has the same value for all units in the population. Another method, the FCI algorithm, is applicable when causal sufficiency does not hold \citep{spirtes1999algorithm}. However, while the PC algorithm can identify direct causes, FCI algorithm cannot distinguish between direct and indirect causes. 

Let $P$ denote the probability distribution of $Y_v, v\in V$. The PC algorithm also assumes \emph{faithfulness} of the DAG $G$ to $Y_v, v\in V$: if the DMP with respect to $G$ encompasses all the conditional independence relations due to $P$, $G$ is said to be \emph{faithful} to $Y_v, v\in V$. Using a consistent statistical test for conditional independence, and assuming causal sufficiency and faithfulness, the PC algorithm estimate, $\hat{G}$, is consistent for $G$; that is, $\hat{G}$ converges in probability to $G$ with increasing number of samples in data \citep{zhang2012strong,spirtes2000causation}.

\section{Causal Functional Connectivity for Time Series}

In this section, we propose a novel methodology for modeling and estimation of the CFC for time series. This methodology is generic and applicable to various time series including neural recordings. The CFC is represented by a graph with nodes as neurons, each corresponds to a time series of neural activity, and edges indicating causal connectivity. In the following, we show we can explicitly model causal relations within and between time series in a DPGM framework and use it to define the CFC.

\subsection{Unrolled Graphical Modeling of Time Series}
\label{subsec: causalfc-temporal}
\begin{figure}[t]
    \centering
    \includegraphics[width = \textwidth]{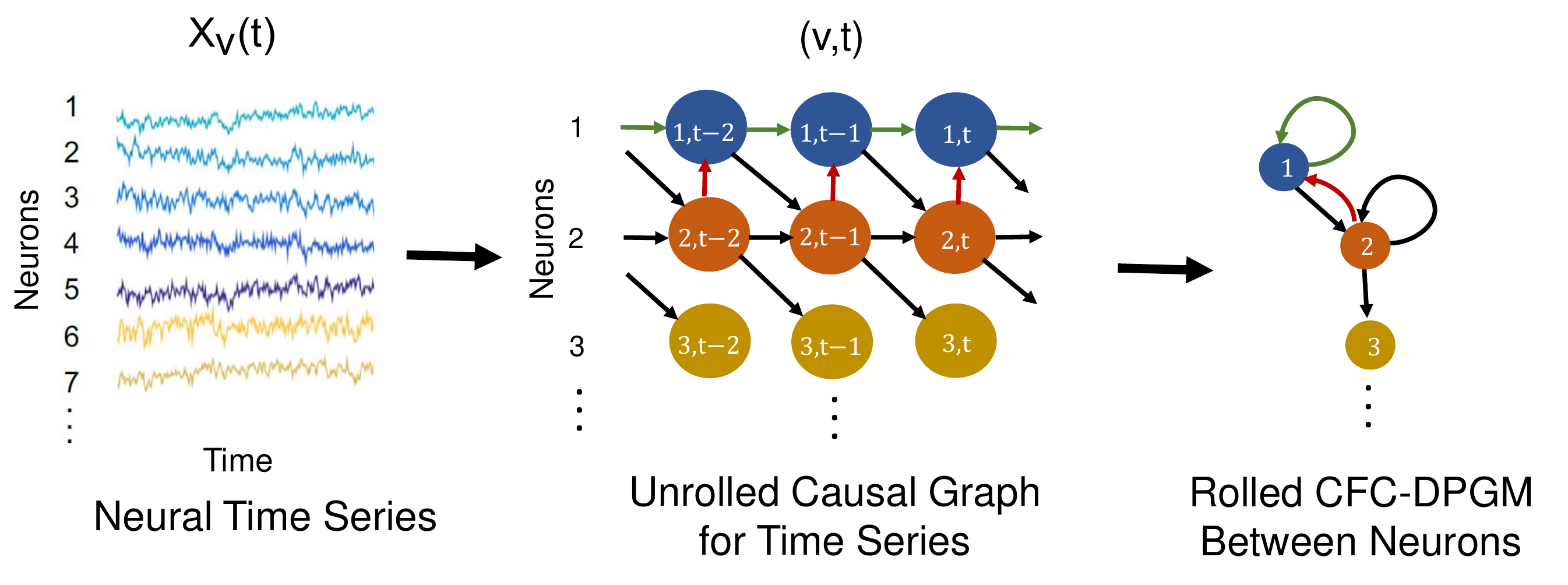}
    
    \caption{Causal modeling of the neural time series (left) by the \emph{Unrolled Causal Graph} (middle), and then rolling back its edges (red) to define the \emph{Rolled CFC-DPGM} (right).}\label{fig:causal-def-example}
\end{figure}

We aim to incorporate the causal influence of the activity of neuron $u$ at time $t_1$ upon the activity of neuron $v$ at time $t_2$ in our proposed model. To do that, we \emph{unroll} the time series $X_v(t), v\in V,0\leq t\leq T$ into nodes $(v,t)$ for neuron $v$ and time $t$ where the node $(v,t)$ corresponds to the variable $X_v(t)$. We use directed edges between the nodes $(v,t)$ to represent causal relations between $X_v(t)$. For example, the edge $(u,t_1)\rightarrow (v,t_2)$ represents the causal influence of the activity of neuron $u$ at time $t_1$, $X_v(t_1)$ upon the activity of neuron $v$ at time $t_2$, $X_v(t_2)$. Let $\bm{V} = \{(v,t), v\in V,0\leq t\leq T\}$ be the set of nodes in the unrolled time series, $\bm{E}$ be the set of directed edges between the nodes and $\bm{G}=(\bm{V},\bm{E})$ be the \emph{Unrolled Graph} for the time series (See Figure \ref{fig:causal-def-example}-middle). We assume that $\bm{G}$ is a DAG in which there are no cycles in the edges in $E$.

Causal relationships between neurons are either forward in time: (1) from neurons $u$ at time $t_1$ to neuron $v$ at time $t_2$ for $t_1<t_2$, represented by $(u,t_1)\rightarrow (v,t_2)$ for $t_1<t_2$ in $\bm{G}$; or, (2) contemporaneous: when causal influences occur more rapidly than the sampling interval of the time series, represented in $\bm{G}$ by $(u,t)\rightarrow (v,t)$ at time $t$ \citep{swanson1997impulse,runge2019inferring}. Causal relations cannot direct backward in time, that is, $\bm{G}$ will not contain $(u,t_1)\rightarrow (v,t_2)$ for $t_1>t_2$. Furthermore, for the contemporaneous causal influences, we do not allow the activity of $u$ at time $t$ to have causal influence on itself at time $t$, that is $(u,t)\rightarrow (u,t)$ is not allowed, while $(u,t)\rightarrow (v,t)$ is allowed. These considerations imply the absence of cycles in $\bm{G}$, thereby justifying the assumption for $\bm{G}$ to be a DAG to model the causal interactions in the unrolled time series.

In practice, the true causal interactions between $X_v(t)$ are unknown. Yet, when $\bm{X} = \{X_v(t):(v,t)\in \bm{V}\}$ satisfies the DMP with respect to DAG $\bm{G}$, then it implies that $\bm{G}$ captures the causal functional interactions among $X_v(t)$, as we show in the functional equivalence of DMP in Section \ref{sec:dgm}. We refer to such a DAG $\bm{G}$ as the \emph{Unrolled Causal Graph} for the time series $X_v(t),v\in V,t\in T$.

\subsection{Rolled CFC-DPGM}
Typically for signals, effective CFC representation refers to relations between neurons rather than relations between different signals' times. Thereby, we propose to roll back the unrolled causal DAG $\bm{G}$ to define CFC between neurons (See Figure \ref{fig:causal-def-example}-right). The rolled graph is based on the principle that the existence of a causal relationship from neuron $u$ at time $t_1$ to neuron $v$ at time $t_2$, $(u,t_1)\rightarrow (v,t_2)\in \bm{G}$ for $t_1<t_2$, would imply that $u$ is connected to $v$ in the rolled CFC. In practice, the causal interactions weaken as the time-gap $t_2-t_1$ grows. Thereby, we consider a maximum time-delay of interaction, $\tau$, so that $(u,t_1)$ and $(v,t_2)$ would not share a significant influence between them if the time gap $t_2-t_1>\tau$ for all neurons $u,v$. Such a consideration of the maximum time-delay aids in making statistical inference from the time series data. Thus, if $(u,t_1)\rightarrow (v,t_2)$ in $\bm{G}$ for some $t_1\leq t_2\leq t_1+\tau$, then the CFC graph between neurons should include $u\rightarrow v$. We consolidate these concepts to define causal functional connectivity between neurons based on DPGM, in the following.

\begin{definition}[Rolled CFC-DPGM]\label{def:causalfc_td} Let $\bm{X}$ satisfy the DMP with respect to DAG $\bm{G}=(\bm{V},\bm{E})$. The \emph{Rolled CFC-DPGM} for neurons in $V$ with maximum time-delay of interaction $\tau$, is defined as the directed graph $F_{\tau}$ having edge $u\rightarrow v$ if $(u,t_1)\rightarrow (v,t_2) \in \bm{E} \text{ for some } 0\leq t_1 \leq t_2 \leq t_1+\tau$. $u$, $v\in V$ could be either the same neuron or distinct neurons.
%$i\rightarrow j \in E_{\tau}$ if $(i,t_1)\rightarrow (j,t_2) \in \bm{E}$ for some $0\leq t_1 \leq t_2 \leq t_1+\tau$, for $i,j \in V$ ($i$, $j$ can be equal).
\end{definition}

We show an example in Figure \ref{fig:causal-def-example}, where for neurons $V=\{1,2,3\}$, their unrolled DAG $\bm{G}$ is represented by Figure \ref{fig:causal-def-example}-middle. When the neural data $\bm{X}$ satisfies the directed Markov Property with respect to $\bm{G}$, the CFC graph with maximum time delay of interaction $\tau = 1$ is given by Figure \ref{fig:causal-def-example}-right. Note that the same CFC graph would be obtained by taking any value of $\tau\geq 1$.

\textbf{Property}: The transformation from unrolled DAG $\bm{G}$ to Rolled CFC, $F_{\tau}$, is a well-defined function, meaning that starting from the same unrolled DAG $\bm{G}$ we will not have multiple possible CFC and there will be a unique CFC $F_{\tau}$.\\ 
\underline{Proof}: By contradiction, consider two distinct CFCs $F_{\tau,1}$ and $F_{\tau,2}$ with $F_{\tau,1}\neq F_{\tau,2}$ obtained from the unrolled DAG $\bm{G} = (\bm{V}, \bm{E})$. Since $F_{\tau,1}\neq F_{\tau,2}$, so $\exists i,j\in V$ such that $i\rightarrow j \in F_{\tau,1}$ but $i\rightarrow j \notin F_{\tau,2}$, where $i,j$ could be either the same or distinct neurons. Using the definiton \ref{def:causalfc_td} of Rolled CFC-DPGM, $i\rightarrow j \in F_{\tau,1}$ implies that for some $0\leq t_1\leq t_2\leq t_1 + \tau$, $(i,t_1)\rightarrow (j,t_2) \in \bm{E}$. But $i\rightarrow j \notin F_{\tau,2}$ contradicts this as it implies that $(i,t_1)\not\rightarrow (j,t_2) \in \bm{E}$ for any $0\leq t_1\leq t_2\leq t_1 + \tau$. 

\paragraph{Contemporaneous and Feedback Interactions} We highlight that the Rolled CFC-DPGM in Definition \ref{def:causalfc_td} incorporates contemporaneous interactions, which can arise if the causal influences occur more rapidly than the sampling interval of the time series or the aggregation interval for aggregated time series. Such a scenario can arise for example in spiking neural datasets where peri-stimulus time histograms aggregate the spike trains over time intervals \citep{shinomoto2010estimating,stevenson2008inferring}, and in Functional Magnetic Resonance Imaging (fMRI) datasets where low sampling rates is typical \citep{stephan2007comparing}. However, if the time span in sampling and aggregation is expected to be less than the time scale of causal interactions, then one can impose $t_1 < t_2$ to exclude contemporaneous interactions. Additionally, the Rolled CFC-DPGM accomodates self-loops in neural interactions \citep{mullins2016unifying,sheffield2016cognition}, by checking whether $(u,t_1)\rightarrow (u,t_2)\in \bm{E}$ for some $0\leq t_1< t_2\leq t_1+\tau$ in determining whether $u\rightarrow u \in F_{\tau}$. Longer feedback loops are also incorporated. For example, the existence of $u\rightarrow v\rightarrow u \in F_{\tau}$ is determined by checking whether  $(u,t_1)\rightarrow (v,t_2)$ and $(v,t_2)\rightarrow (u,t_3)\in \bm{E}$ for some $0\leq t_1\leq t_2\leq t_3\leq t_1+\tau$. By virtue of the technique of unrolling and rolling back, the Rolled CFC-DPGM captures causality while including cycles, since the Unrolled Graph is still a DAG and thereby meets the requirement for satisfying DMP.

 \begin{figure}[t!]
    \centering
    \vspace{5mm}
    \includegraphics[width = \textwidth]{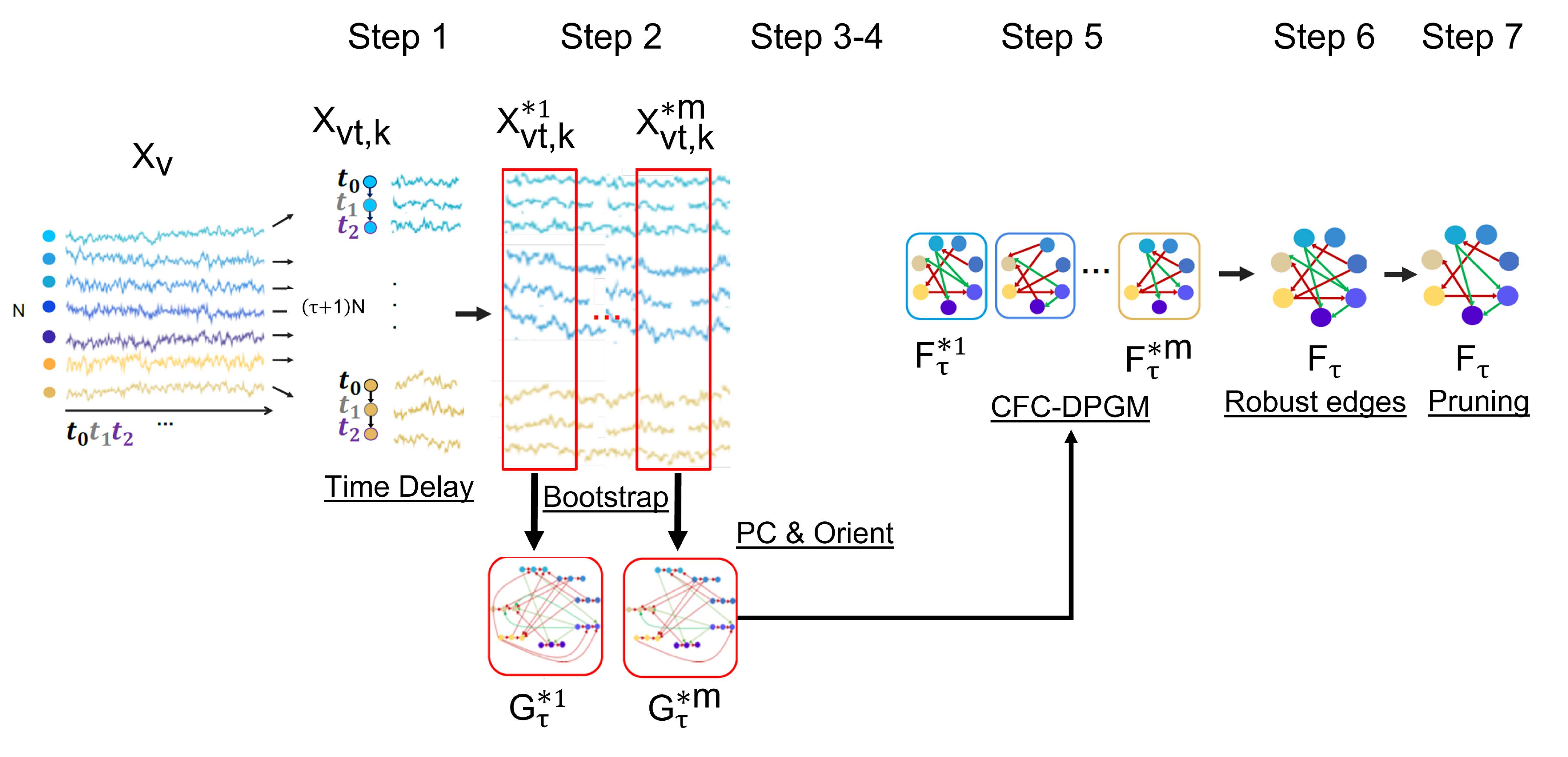}
    \caption{Illustration of Steps 1-7 in the TPC Algorithm: Time Delay, Bootstrap, PC, Orient, Rolled CFC-DPGM, Robust edges and Pruning.}
    \label{fig:TPC}
\end{figure}

\subsection{Estimation from data: Time-Aware PC (TPC) Algorithm} \label{alg:estimate-cfcpdgm}

In this section we outline the steps to estimate the Rolled CFC-DPGM from dataset $\bm{X}=\{X_v(t):v\in V, 0\leq t\leq T\}$, $\dim V=N$, which constitute the Time-Aware PC (TPC) Algorithm. The output of TPC is a Rolled CFC that contains edges found significant in the duration of recording. The TPC Algorithm first aims to estimate the unrolled DAG for $X_v(t), v\in V, 0\leq t \leq \tau$, considering maximum time-delay of interaction $\tau$. The unrolled DAG has nodes as $(v,t), v\in V, 0\leq t \leq \tau$ and with respect to which $X_v(t), v\in V, 0\leq t \leq \tau$ satisfies the DMP. To perform the estimation, TPC constructs samples for a node $(v,t)$ consisting of time delayed instances of $X_v(t)$, denoted $X_{vt,k}$, with a time delay of $2(\tau+1)k$, i.e. with a shift backward of the signal $X_v$ by $2(\tau+1)k$. Such a time-delay of multiples of $2(\tau+1)$ ensures a substantial time-gap of $2(\tau+1)$ units between samples, that reduces interdependence between the samples, considering the maximum time-delay of interaction $\tau$. This operation increases the number of considered time-series by a multiple of $\tau+1$ such that for original recordings of $N$ nodes in $X_v$, there would be $N(\tau+1)$ nodes in $X_{vt}$.

\begin{algorithm}[ht!]
\caption{TPC\label{algo:TPC}}
\SetAlgoLined
\SetKwInOut{Input}{Input}\SetKwInOut{Output}{Output}
\Input{Recordings of activity of neurons in $V$ over time $(X_v(t):v\in V, t\in 0:T)$; maximum delay of interaction between neurons $\tau$; significance level $\alpha$;\\ 
For bootstrapping: window width $L$; $m$ iterations; bootstrap stability cutoff $\gamma$;}
\Output{CFC estimate, denoted $F_{\tau}$.}
\Begin{
\begin{enumerate}
    \item[Step 1.] \underline{Time-Delayed Samples.} For $v\in V, 0\leq t\leq \tau$, construct $k$-th sample for $(v,t)$, denoted  $X_{vt,k}$, by time-delay of $2(\tau+1)k$: $X_{vt,k}=X_v(t+2(\tau+1)k)$, $0\leq k \leq K = \frac{T-\tau}{2(\tau+1)}$.
    \item[Step 2.] \underline{Bootstrap.} For $v\in V, 0\leq t\leq \tau$, select a random window from $X_{vt,k}$ to obtain $X_{vt,k}^*$, by drawing a random integer $r\in [0,K-L]$, and $X_{vt,k}^* = X_{vt,r+k}, 0\leq k \leq L$.
    \item[Step 3.] \underline{PC.} Use PC algorithm to estimate the unrolled DAG for $X_v(t)$ over $v\in V$, $0\leq t\leq \tau$, with the samples $X_{vt,k}^*, 0\leq k \leq L$, for $X_v(t)$. Denote the output as $\bm{G}_{\tau}^*$.
    \item[Step 4.] \underline{Orient.} Reverse the edge directions of edges $(v,t_1)\rightarrow (u,t_2)\in \bm{G}_{\tau}^*$, when $0\leq t_2<t_1 \leq \tau, u\in V, v\in V$ and update $\bm{G}_{\tau}^*$. 
    \item[Step 5.]\underline{Rolled CFC-DPGM.} \begin{enumerate}\item Convert $\bm{G}_{\tau}^*$ to rolled CFC-DPGM $F_{\tau}^*$ with max. delay of interaction $\tau$ using Def. \ref{def:causalfc_td}.
    \item Find the connectivity weights $w_{\tau}^{*}(u,v)$ for connections $u\rightarrow v \in F_{\tau}^{*}$ by Def. \ref{def:causalFCweight}.
    \end{enumerate}
    \item[Step 6.]\underline{Robust edges.} \begin{enumerate} \item Repeat Steps 2-5 to obtain $m$ iterates of the rolled CFC-DPGM: $F_{\tau}^{*i}$, and connectivity weights $w_{\tau}^{*i}(u,v)$, $i=1:m$.
    \item Output a single CFC, $F_{\tau}$, with only those edges whose relative frequency of occurrence among $F_{\tau}^{*i}$ is above $\gamma$.
    \item Output single Connectivity Weights, $w_{\tau}(u,v)$, for connections $u\rightarrow v\in F_{\tau}$, as the average of $\{w_{\tau}^{*i}(u,v):u\rightarrow v \in F_{\tau},u\rightarrow v \in F_{\tau}^{*i}, i\in 1:m\}$ when the set is non-empty and $0$ otherwise.
    \end{enumerate}
    \item[Step 7.]\underline{Pruning.} Remove from $F_{\tau}$ those edges $u\rightarrow v \in F_{\tau}$ with $\vert w_{\tau}(u,v) \vert < w_0$, where, $w_0 = \frac{1}{10}\max\{ \vert w_{\tau}(u,v) \vert: u\rightarrow v \in F_{\tau}\}$.
\end{enumerate}
}
\end{algorithm}

In Step 2, TPC selects random windows from $X_{vt,k}$ to obtain $X_{vt,k}^*$. The process is called \emph{bootstrap} since in the next step, on each window, the PC algorithm is applied to estimate the unrolled DAG for $X_v(t), v\in V, 0\leq t \leq \tau$, with nodes being $(v,t), v\in V, 0\leq t\leq \tau$. These operations provide a set of graphs modeling a sample of the unrolled DAG of each window. 

In Step 3, on each window, the PC algorithm outputs the unrolled DAG as a completed partially directed acyclic graph (CPDAG), defined as the graph union of DAGs that satisfy DMP with respect to the window, and denoted by $\bm{G}_{\tau}^*$ \citep{pearl2009causal}. 

In Step 4, TPC \emph{corrects} the edges in $\bm{G}_{\tau}^*$ which direct from future to past time by reversing them, to be consistent with the temporal direction of causal interactions in the time series.

In Step 5, the re-oriented Unrolled graph $\bm{G}_{\tau}^*$ is transformed to give the Rolled CFC-DPGM denoted $F_{\tau}^*$. At this step, weights for edges in $F_{\tau}^*$ are also obtained using interventional connectivity weights (\ref{def:causalFCweight}), that quantifies the causal effect of intervention on each neuron to its connected neurons.

In Step 6, a single CFC consensus $F_{\tau}$ is obtained by keeping only those edges which have relative frequency of occurrence among $F_{\tau}^{*i}$ to be greater than the cut-off $\gamma$, where $F_{\tau}^{*i}, i=1,\ldots,m$ is obtained over $m$ iterations of Steps 2-5. A single connectivity weight consensus for an edge in $F_{\tau}$ is achieved by averaging over the weights for the same edge, whenever present, over the $m$ iterates. The resampling procedure promotes detection of stable edges \citep{maathuis2009estimating}. 

Finally, in Step 7, $F_{\tau}$ is pruned to further reduce spurious edges, by removing the edges which have exceedingly low connectivity weights, determined by those edges in $F_{\tau}$ whose weights are less than a tenth in magnitude compared to the maximum magnitude for edge weights in $F_{\tau}$. The TPC algorithm is outlined in Alg. \ref{algo:TPC} and Figure \ref{fig:TPC}.

\subsection{Connectivity weights in the CFC}
In this section, we define the connectivity weights obtained by TPC. Connectivity weights refer to a weight $w_{uv}$ for connections $u\rightarrow v$ in the CFC graph. We consider \emph{interventional causal effects} to define connectivity weights. Interventional causal effects quantify how much effect an intervention applied to neuron $u$ will have on neuron $v$.

\paragraph{Definition 2: Interventional causal effects in Unrolled DAG.}
Let $\bm{X}$ satisfy the Directed Markov Property with respect to $\bm{G}=(\bm{V},\bm{E})$. The \emph{interventional causal effect} of $X_u (t) = x_{u,t}$ on $X_v (t')$, where $x_{u,t}$ are fixed values, for $u,v\in V, 0\leq t\neq t'\leq T$,  is defined in interventional calculus by Pearl et. al as follows \citep{pearl2009causality,huang2012pearl,maathuis2009estimating} 
\begin{equation}\label{eq:causaleffect}
\frac{\partial}{\partial x} E(X_v (t') \vert X_u (t) = x) \vert_{x=x_{u,t}}\end{equation} Assuming $X_u (t), u\in V$ are jointly Gaussian, the causal effect does not depend on the value of $x_{u, t}$, and the causal effect of $X_{u} (t)$ on $X_v (t')$ from Eq. (\ref{eq:causaleffect}) takes the following form,
\begin{equation}\label{eq:causaleffectgaussian}
    w_{u,t}^{v,t'}=\left\{\begin{array}{lr}
    0, & \text{ if }(v,t') \in pa_{\bm{G}}(u,t),\\
    \text{coefficient of }X_u (t) \text{ in }X_v (t') \sim X_u (t) + X_{pa_{\bm{G}}(u,t)} & \text{ if } (v,t') \notin pa_{\bm{G}}(u,t)
    \end{array}\right.
\end{equation}
where $X_v (t') \sim X_u (t) + X_{pa_{\bm{G}}(u,t)}$ is shorthand for linear regression of $X_v (t')$ on $X_u (t)$ and $X_{pa_{\bm{G}}(u,t)} = \{X_a (b) : (a,b) \in pa_G(u,t)\}$.

Note that when $\bm{X}$ satisfy the Directed Markov Property with respect to $\bm{G}=(\bm{V},\bm{E})$, the causal effects are defined in interventional calculus literature for all pairs of nodes in $\bm{V}$ and not that only for those pairs which are adjacent. And, if $u\rightarrow v \in \bm{E}$ then the interventional causal effect from $v$ to $u$ is $0$ \citep{maathuis2009estimating}.

Under the Gaussian assumption, we define the \emph{interventional causal effect} of the activity of neuron $u$ at time $t$ on the activity of neuron $v$ at time $t'$ to be $w_{u,t}^{v,t'}$ for $0\leq t<t'\leq T$, $u,v \in V$. Using this, we define weights for the connection from $u$ to $v$ for $u,v$ in the rolled CFC-DPGM $F_{\tau}$, following the way $F_{\tau}$ is defined from $\bm{G}$.
%\begin{definition}The weight of $(i,k)\rightarrow (j,l)$ in $\bm{G}$ is defined by $w_{i,k}^{j,l}$ in Equation (Eq. \ref{eq:causaleffectgaussian}). \end{definition}
\paragraph{Definition 3: Interventional connectivity weights in Rolled CFC-DPGM.}\label{def:causalFCweight}
Let $\bm{X}$ satisfy DMP with respect to the DAG $\bm{G}=(\bm{V},\bm{E})$ and $F_{\tau}$ is the Rolled CFC-DPGM with max delay $\tau$. If neurons $u,v$ are connected as $u\rightarrow v$ in $F_{\tau}$, then, the \emph{weight of connection} from neuron $u$ to $v$ with max delay $\tau$, denoted by $w_{\tau}(u,v)$, is defined as the average of the causal effects: $w_{u,t}^{v,t'}$ for $(u,t)\rightarrow (v,t')\in \bm{E}$, $0\leq t\leq t'\leq t+\tau$.

\paragraph{Connectivity weights in TPC Algorithm.} After the CFC graph $F_{\tau}^*$ is obtained in Step 3-5a in TPC algorithm, the interventional connectivity weights for connections in $F_{\tau}^*$ are obtained in Step 5b to define the connectivity weights $w_{\tau}^{*i}(u,v)$ for connections $u\rightarrow v \in F_{\tau}^{*i}$. Then bootstrapping in Step 6 ensures greater stability of the estimated connectivity weights. Step 6 outputs a single connectivity weight $w_{\tau}(u,v)$ for connections $u\rightarrow v$ in $F_{\tau}$, as the average of $\{w_{\tau}^{*i}(u,v):u\rightarrow v \in F_{\tau},u\rightarrow v \in F_{\tau}^{*i}, i\in 1:m\}$ when the set is non-empty and $0$ otherwise. Therefore, this finds a connectivity weight for the edge $u\rightarrow v\in F_{\tau}$ by taking the average of connectivity weight of the edge $u\rightarrow v \in F_{\tau}^{*i}$ whenever it exists over $i$.

\paragraph{Pruning by Connectivity Weights.} After the rolled CFC-DPGM and Connectivity Weights have been inferred by the TPC Algorithm, spurious connections can be pruned further in Step 7 of TPC (\ref{algo:TPC}), by discarding those connections whose connectivity weight is less than a threshold. For this threshold, we use a factor of 10 of the maximum Connectivity Weight in the rolled CFC-DPGM.

From the interpretation of regression coefficients in Eq. (\ref{eq:causaleffectgaussian}), a negative connectivity weight (\ref{def:causalFCweight}) from neuron $u\rightarrow v$ in $F_{\tau}$ indicates an inhibitory connection, in which, increased activity $X_u (t)$ of the pre-synaptic neuron $u$ at time $t$ causes subjugation of activity $X_v (t')$ of the post-synaptic neuron $v$ at a following time $t'$, when activity of the neurons that are causally connected to neuron $v$ at time $t$, are kept fixed. In a similar manner, a positive FC weight from neuron $u\rightarrow v$ in $F_{\tau}$ indicates an excitatory connection. In this way the strength of the functional connection, which also indicates it's excitatory and inhibitory nature is learnt from the data. 
 
\subsection{Properties of Rolled CFC-DPGM}\label{sec:properties}
We highlight properties of the Rolled CFC-DPGM, obtained by TPC, in capturing causal relationships in neural dynamics in a non-parametric manner and being predictive of the impact of counterfactual interventions to the neurons in the CFC.

\paragraph{Non-parametric Causal Relations.} The following theorem shows that the CFC given by the model is consistent with the ground truth causal relationships between neural activity at different time without requiring any assumptions on the functional form of the relationships. That is, we show that if past time points of neurons in $A_v\subset V$ influence the present time point of $v\in V$ by an arbitrary function with independent random noise, then neurons in $A_v$ are connected to the neuron $v$ in their Rolled CFC-DPGM. This means that causal relationships among the neurons, in their unknown arbitrary dynamical equation, are accurately represented by the CFC without prior knowledge of the functional form of the relationships. The benefit of Rolled CFC-DPGM is that it uses a non-parametric graphical model for the temporal relationships between neurons and does not assume a parametric equation for the temporal relationships. Furthermore, the Rolled CFC-DPGM provides a framework to answer causal questions related to the consequence of interventions and counterfactuals.

\begin{theorem}[Consistency with Time Series Causal Relations]\label{thm:causal-fc-td}
For neurons $v\in V$ with activity $X_v(t)$ at time $t$, if
\begin{equation}\label{eq:thmfirst}
  X_v(t) = g_{v,t} (X_{u_{v,1}}(t_{v,1}),\ldots,X_{u_{v,K}}(t_{v,K}),\epsilon_{v}(t)),
\end{equation}
$\text{for some } u_{v,i} \in V$ and $t_{v,i}\in [t-\tau,t]$ with either $u_{v,i}\neq v$ or $t_{v,i}\neq t$, for $1\leq i\leq K$, where $\tau$ is the maximum time-delay of interaction, and $g_{v,t}$ is a measurable function and $\epsilon_v(t)$ are independent random variables, then the graph $F_{\tau}$ with nodes $V$ and parents of $v$, $pa_{F{\tau}}(v)$, given by \[pa_{F{\tau}}(v) = \{u_{v,1},\ldots, u_{v,K}\}\] is the Rolled CFC-DPGM between the neurons in $V$.
\end{theorem}
\vspace{-4mm}
\begin{proof}
See Appendix \ref{proof:thm1}.
\end{proof}

\paragraph{Interventional Properties} The Rolled CFC-DPGM can answer questions concerning counterfactual interventions on neurons without experimentally performing the interventions, i) Ablation of a neuron $A$, ii) Activity of neuron $B$ is externally modulated. In the following corollary we show how conclusions can be drawn for such queries. 

\begin{corthm}[Intervention]\label{cor:intervention}
For neurons $v\in V$ following the dynamics in equation Eq. (\ref{eq:thmfirst}), let us consider there is an experimental or counterfactual intervention on neurons $v_1,\ldots,v_k \in V$ during $t\in T_I$, such as ablation or external control. 1) For ablation of $v_1,\ldots, v_k$ the connections incident as well as outgoing from them are removed. 2) During $t\in T_I$, for external control, all connections incident on $v_1,\ldots, v_k$ are removed in the Rolled CFC-DPGM $F_{\tau}$ and other connections remain intact. 
\end{corthm}

\vspace{-4mm}
\begin{proof}
See Appendix \ref{proof:thm2}.
\end{proof}

\vspace{-2mm}
This corollary justifies the usage of causal reasoning with the edges of the CFC alone to answer the interventional queries. Answers to the questions by causally reasoning are as follows: 

i) When neuron $A$ is ablated, one just deletes all the edges incident and originating from neuron $A$ since $A$ has a fixed value after ablation and neither do other neurons influence the activity of neuron $A$ nor does $A$ influence the activity of any other neuron. 

ii) When activity of neuron $B$ is externally controlled, one simply removes the edges incident on neuron $B$, because activity of neuron $B$ no longer depends on its parent neurons in the CFC obtained before intervention rather the activity of neuron $B$ depends on the external control. Edges originating from neuron $B$ in the CFC from before the intervention should remain intact during the intervention since the functional pathways from $B$ to its descendant neurons in the CFC remain intact during external control. To illustrate the properties of non-parametric causal relations (Theorem \ref{thm:causal-fc-td}) and interventions (Corollary \ref{cor:intervention}), we consider the following example.

\begin{figure}[t]
    \centering
    \includegraphics[width=0.7\textwidth]{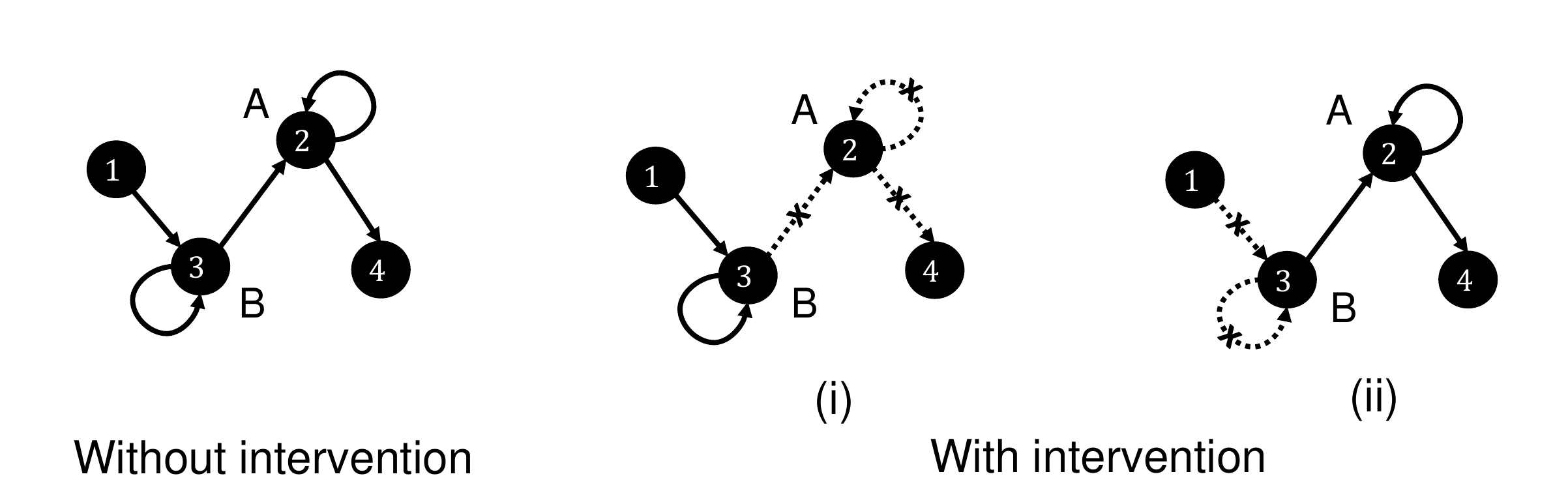}
    \caption{Rolled CFC-DPGM (left) for neurons 1-4 with dynamics as in Example \ref{ex:dir-gm}, and consequence of intervention on neurons labelled A and B by (i) Ablation of A and (ii) External modulation of B.}
    \label{fig:example4}
\end{figure}

\begin{exmp}\label{ex:dir-gm} Let $V$ denote a network of 4 neurons, labeled $\{1,2,3,4\}$ with neural activity $X_v(t), v\in V$ related as, 
\begin{align*}
    X_1 (t) &= g_{1,t} (\epsilon_{1}(t))\\
    X_2 (t) &= g_{2,t} (X_2 (t-1), X_3(t-1), X_3(t-2), \epsilon_{2}(t))\\
    X_3 (t) &= g_{3,t} (X_3 (t-1), X_1(t-1), \epsilon_{3}(t))\\
    X_4 (t) &= g_{4,t} (X_2(t-1), \epsilon_{4}(t))
\end{align*}
for independent random variables $\epsilon_v (t)$ and measurable functions $g_{v,t}$, $v=1,2,3,4; 0\leq t\leq 1000$ msec. By Theorem \ref{thm:causal-fc-td} it follows that the graph: $2\rightarrow 2, 3\rightarrow2, 3\rightarrow 3,1\rightarrow 3, 2\rightarrow 4$ as in Figure \ref{fig:example4}-left is the CFC among the neurons considering maximum time delay of interaction to be 1 msec or higher. Suppose one asks the question of type (i), how would the functional connectome change if neuron (A) were ablated? According to Corollary \ref{cor:intervention}, the resulting CFC would be $1\rightarrow 3, 3\rightarrow 3$ as in Figure \ref{fig:example4}-middle by removing the connections to and from neuron $2$ according to Corollary \ref{cor:intervention}. Suppose one asks the question of type (ii), how would the functional connectome change if activity of neuron (B) were to be externally controlled by optogenetics? According to Corollary \ref{cor:intervention}, the resulting CFC would be $3\rightarrow 2, 2\rightarrow 2, 2\rightarrow 4$ as in Figure \ref{fig:example4}-right by removing the parent connections of neuron $3$ according to Corollary \ref{cor:intervention}. 
\end{exmp}

\section{Comparison Study of TPC with other Approaches to Causal Functional Connectivity}\label{sec:valid}
We compare the performance of TPC with different existing CFC inference approaches to recover relationships in ground truth dynamical equations by generating synthetic data from three simulation paradigms. In particular, we estimate their CFC using GC, DPGM and TPC. The simulation paradigms correspond to specific model assumptions to assess the impact of model assumptions on the performance of the approaches (See Appendix \ref{simul:details}).

We measured the algorithms' performance using CFC inference for 25 different simulations and summarized the results using three metrics: (1) Combined Score (CS), (2) True Positive Rate (TPR), (3) 1 - False Positive Rate (IFPR). Let True Positive (TP) represent the number of correctly identified edges, True Negative (TN) represent the number of correctly identified missing edges, False Positive (FP) represent the number of incorrectly identified edges, and False Negative (FN) represent the number of incorrectly identified missing edges across simulations. IFPR is defined as: $\text{IFPR}=\left(1-\frac{\text{FP}}{\text{FP+TN}}\right)\cdot 100,$ which measures the ratio of the number of correctly identified missing edges by the algorithm to the total number of true missing edges. Note that the rate is reported such that $100\%$ corresponds to no falsely detected edges. TPR is defined $\text{TPR}=\left(\frac{\text{TP}}{\text{TP} + \text{FP}}\right)\cdot 100$ as the ratio of the number of correctly identified edges by the algorithm to the total number of true edges in percent. The Combined Score (CS) is given by Youden's Index \citep{vsimundic2009measures,hilden1996regret}, as follows, $\text{CS} = \text{TPR} - \text{FPR}$.

\begin{figure}
    \centering
    \includegraphics[width=0.95\textwidth]{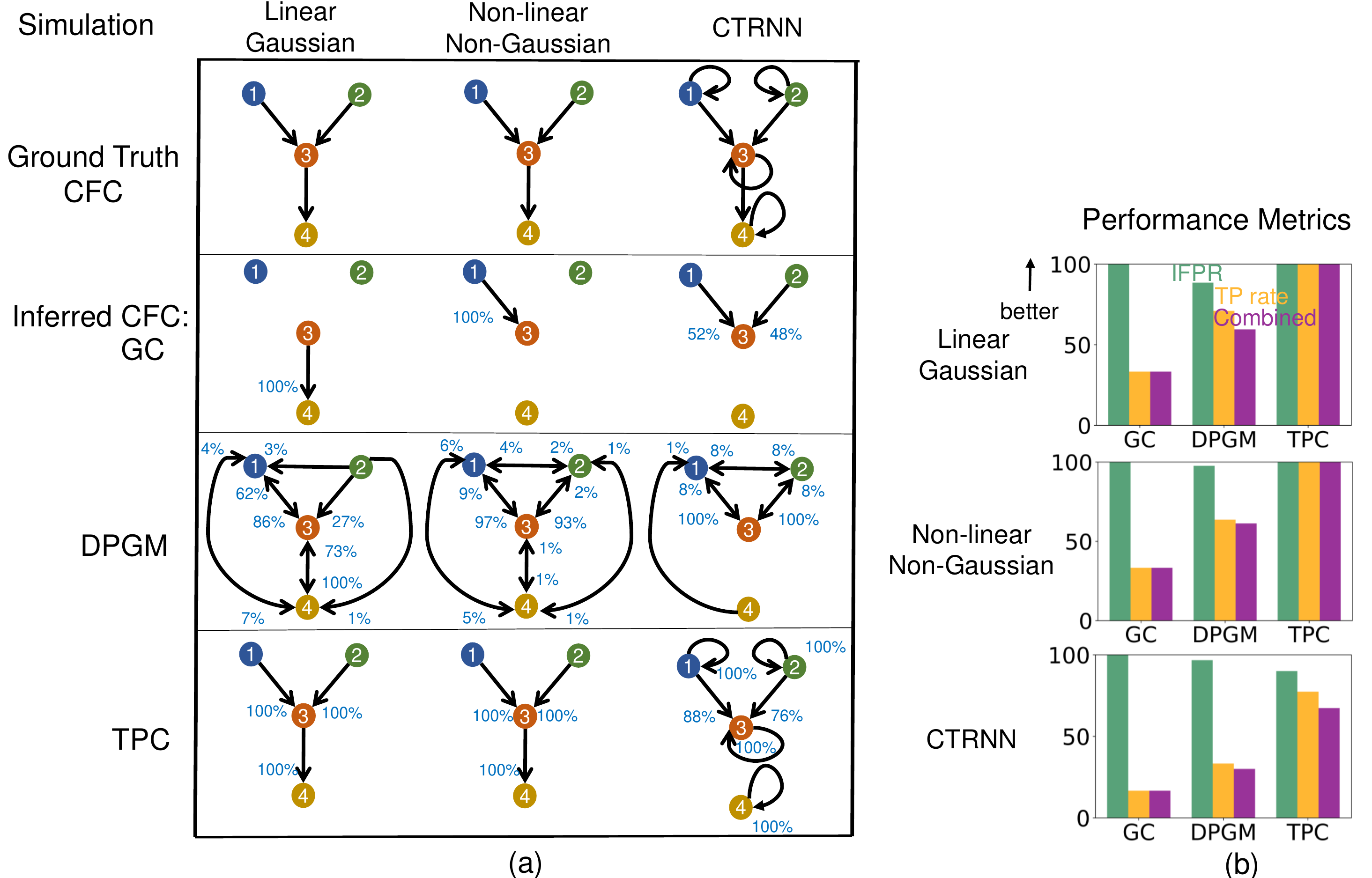}
    \caption{(a) CFC inference by GC, DPGM, and TPC, is compared on three examples of motifs and simulation paradigms; from left to right: Linear Gaussian, Non-linear Non-Gaussian, CTRNN. Table: 4-neurons motifs that define the Ground Truth CFC (row 1) are depicted along  with inferred CFC over several simulation instances according to the three different methods (row 2-4). Each inferred CFC has an edge $v\rightarrow w$ that corresponds to an edge detected in any of the inference instances. The percentage (blue) next to each edge indicates the number of times the edge was detected out of all instances. (b) $\text{IFPR}$ (green) , TP rate (orange) and Combined Score (Purple) of each method are shown for each motif.}
    \label{fig:simul_eval}
\end{figure}

\begin{figure}[t!]
    \centering
    \includegraphics[width=0.8\textwidth]{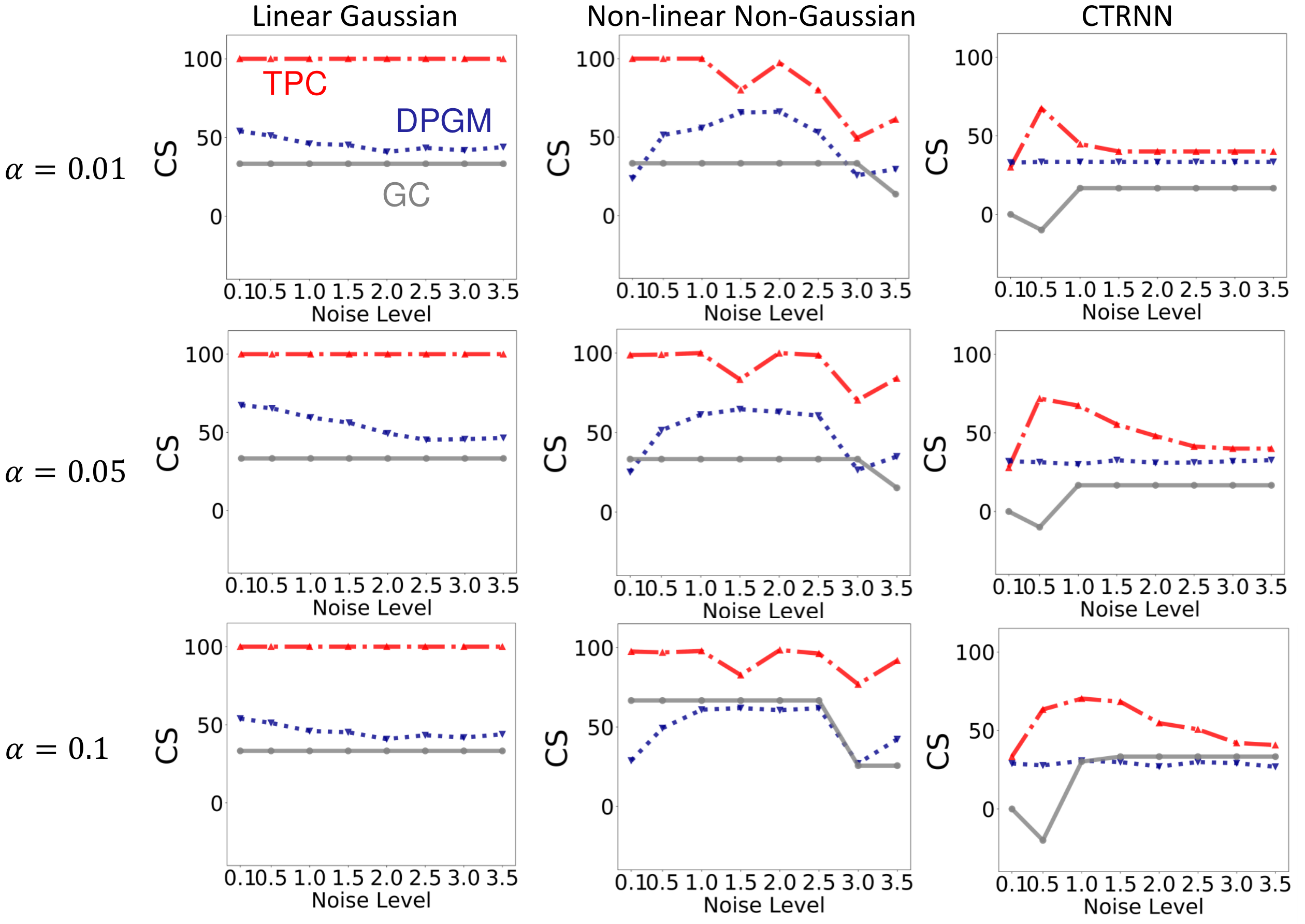}
    \caption{Combined Score of the three methods of CFC inference - TPC (red), DPGM (blue), GC (gray), over varying noise levels in simulation $\eta = 0.1,0.5,1.0,\ldots,3.5$, for simulated motifs from Linear Gaussian, Non-linear Non-Gaussian and CTRNN paradigms (left to right), with thresholding parameter $\alpha = 0.01, 0.05, 0.1$ (top to bottom).}
    \label{fig:comp_noise}
\end{figure}

In the motifs and simulation paradigms that we consider, there are $4$ neurons and $16$ possible edges (including self-loops) per simulation resulting with total of $400$ possible edges across $25$ simulations. Figure \ref{fig:simul_eval} compares in detail the results for GC, DPGM and TPC in inference of true CFC for noise level $\eta =1$ and thresholding parameter $\alpha = 0.05$. Here we also report the percentage of the simulations that has each estimated edge present. Higher percentage indicates higher confidence in the detection of that edge. Figure \ref{fig:comp_noise} compares the Combined Score of the approaches over different values of noise level $\eta$ and thresholding parameter $\alpha$ for each simulation paradigm.

\begin{itemize}[leftmargin=0pt]
    \item[] In \textit{Linear Gaussian scenario (left column in Figure \ref{fig:simul_eval})}, the connections between neurons in the Ground Truth CFC are excitatory due to positive coefficients in the linear dynamical equation for neural activity. GC generates a sparse set of edges in which it correctly detects a single edge $3\rightarrow 4$ among the three edges of the true CFC but misses two other edges. DPGM generates a large number of edges (9 out of 16), many of which are spurious, though it has a high percentage for expected edges in the Ground Truth CFC ($1\rightarrow 3, 3\rightarrow 4$ with $87 \%$ and $100 \%$ respectively). TPC obtains the Ground Truth CFC, with no spurious edges and obtains the expected edges in all of the trials ($1\rightarrow 3, 2\rightarrow 3, 3\rightarrow 4$ with $100 \%$, $100 \%$ and $100 \%$ respectively). Overall, GC, DPGM and TPC produce $\text{IFPR} = 100\%, 88.5\%, 100\%$, $\text{TPR}=33.3\%, 71.0\%$, and $100 \% $, and CS $= 33\%, 59\%, 100\%$ respectively. Thereby, among the three methods, we conclude that TPC detects the edges perfectly, while GC is highly specific to correct edges, but since it does not detect two out of three edges it is not as sensitive as DPGM. 
    \item[] In the \textit{Non-linear Non-Gaussian scenario (second column)}, in the Ground Truth CFC consists of $1\rightarrow 3, 3\rightarrow 4$ excitatory due to $\sin(x)$ being an increasing function, while $2\rightarrow 3$ is an inhibitory connection due to $\cos(x)$ being a decreasing function for $x\in [0,1]$ in the dynamical equation. As previously, GC consistently detects a sparse set of edges (single edge $1\rightarrow 3$ with $100\%$) which is one of the three true edges. DPGM again generates a large number of edges, some of which are spurious. In the majority of trials ($97 \%$ and $93 \%$, respectively), DPGM correctly obtains two of the three true edges $1\rightarrow 3$ and $2\rightarrow 3$. In contrast, TPC obtains no spurious edges and the true edges were detected for all the trials ($1\rightarrow 3$, $2\rightarrow 3$,$3\rightarrow 4$ with $100\%, 100\%, 100\%$). In summary, GC, DPGM and TPC yielded $\text{IFPR} = 100\%, 97.6\%, 100\%$ and $\text{TPR} = 33.3\%, 63.7\%, 100\%$ and CS $=33\%, 61\%, 100\%$. For this scenario, TPC again has the highest performance among the methods. 
    \item[] In \textit{CTRNN scenario (third column)}, self-loops are present for each neuron, and due to positive weights and increasing activation function $\sigma(\cdot)$ in their dynamical equation, the connections in the Ground Truth CFC are excitatory. GC obtains two of the three true non-self edges $1\rightarrow 3, 2\rightarrow 3$  for $52\%, 48\%$ of the trials. DPGM detects spurious edges, but also infers the non-self true edges $1\rightarrow 3, 2\rightarrow 3$ for $100\%$ of the trials. In comparison, TPC infers no spurious edges and all the self true edges for $100\%$ of the trials and non-self true edges $1\rightarrow 3$ and $2\rightarrow 3$ for $88\%, 76\%$ of the trials. In summary, IFPR of GC, DPGM and TPC is $100\%, 96.7\%, 90\%$ and $\text{TPR}$ is $16.7\%, 33.3\%, 77.3\%$ and CS is $17\%, 30\%, 67\%$ respectively. Among all methods, TPC has the highest TPR, followed by DPGM and lastly GC. Since GC does not detect any false edges, it has the highest IFPR, followed by DPGM and lastly TPC, though all of them have IFPR of at least $90\%$. In terms of the CS, TPC has the highest performance compared to other methods.
\end{itemize}

We compare Combined Score of TPC and other approaches across varying levels of simulation noise $\eta$ from $0.1$ to $3.5$ and thresholding parameter $\alpha = 0.01,0.05,0.1$ in Figure \ref{fig:comp_noise}. In the Linear Gaussian scenario, we note that TPC has a CS of $\approx 100\%$ across all levels of simulation noise and thresholding parameter $\alpha$, and is followed by DPGM in performance and lastly GC. In the Non-linear Non-Gaussian scenario, TPC has the highest CS compared to other methods across levels of noise and $\alpha$. In the CTRNN scenario, the performance of all the three approaches is lower compared to the other simulation paradigms for different level of $\eta$ and $\alpha$, yet TPC has higher CS compared to the other methods over the different parameter values.

\begin{figure}[t!]
    \centering
    \includegraphics[width=0.7\textwidth]{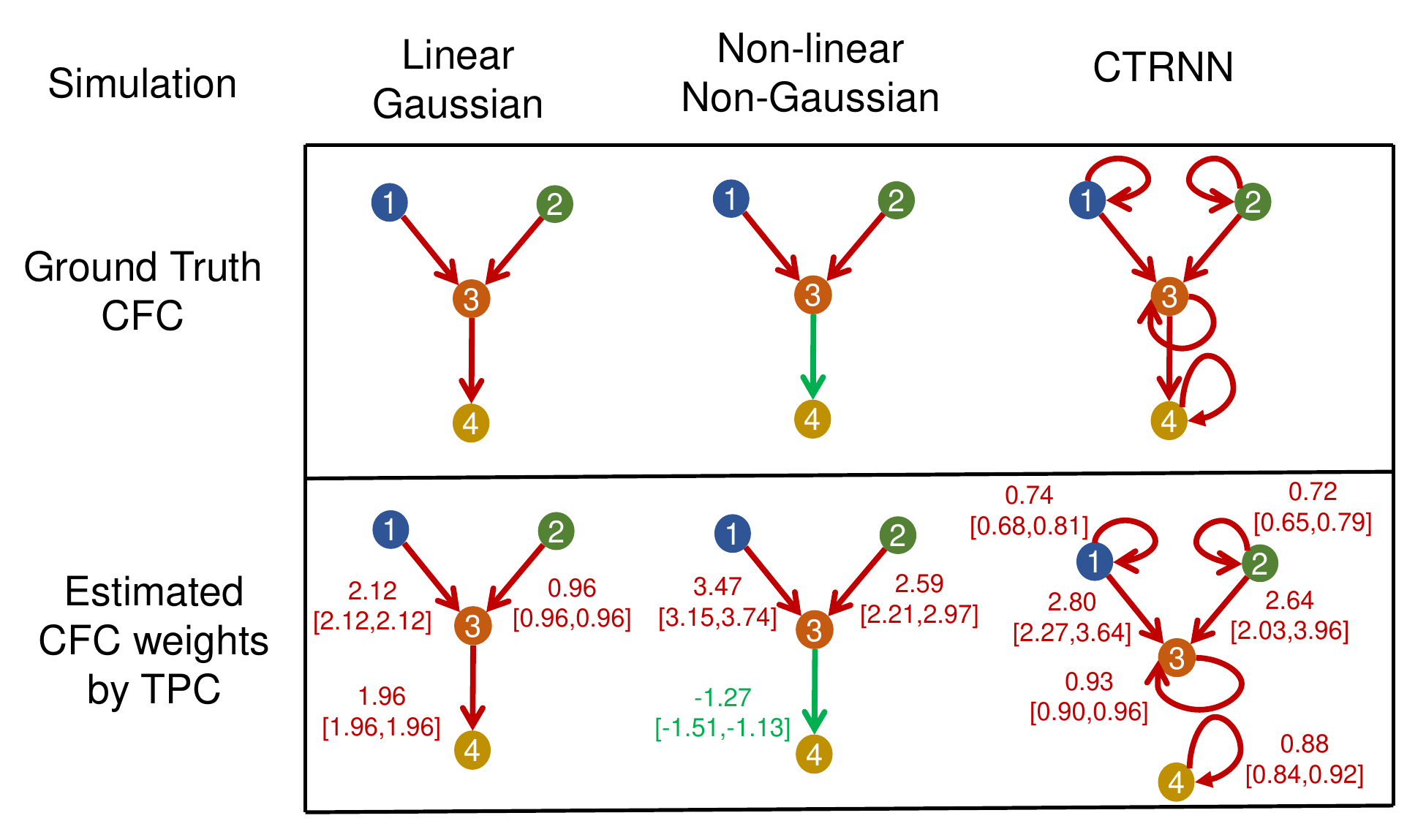}
    \caption{Inference of interventional connectivity weights by the TPC algorithm with max delay $1$ msec for the example motifs from the three simulation paradigms: Linear Gaussian VAR, Non-linear Non-Gaussian VAR, CTRNN (left to right). Top row: Ground Truth CFC with excitatory (red) and inhibitory (green) connections; Bottom row: Estimated CFC labeled with edge weights (median [min,max] over all instances) and inferred nature whether excitatory (red) or inhibitory (green).}
    \label{fig:connweights}
\end{figure}

We demonstrate the connectivity weights obtained by TPC and inferred nature of connections, whether excitatory or inhibitory, across simulations for noise level $\eta=1$ and thresholding parameter $\alpha = 0.05$ in Figure \ref{fig:connweights}. In the \textit{Linear Gaussian scenario}, the estimated connectivity weight of $1\rightarrow 3$, $2\rightarrow 3$, $3\rightarrow 4$ are $2.12, 0.96, 1.96$, across simulation trials. Since the weights are positive, thereby the connections are labeled to be excitatory in all simulation trials, which agrees with the Ground Truth. In the \textit{Non-linear Non-Gaussian scenario}, the estimated weight for the connection $2\rightarrow 3$ is $-1.27$ in median and ranges between $-1.51,-1.13$. Therefore the weight is always negative and labeled inhibitory in the simulation trials. The weight for $1\rightarrow 3,2\rightarrow 3$ are $3.47,2.57$ in median and ranges between $3.15,3.74$ and $2.21,2.97$ respectively. Their weights are always positive and labeled excitatory in all the simulation trials. These labels for the nature of connections obtained by TPC agrees with the ground truth. In the \textit{CTRNN scenario}, the estimated connectivity weight for $1\rightarrow 1, 1\rightarrow 3, 2\rightarrow 2, 2\rightarrow 3, 3\rightarrow 3, 4\rightarrow 4$ are $0.74, 2.80, 0.72, 2.64, 0.93, 0.88$ in median respectively and ranges over positive values in all the simulation trials. Thereby the connections are labeled excitatory in all the simulations trials, which agrees with the Ground Truth.

\section{Application to Benchmark Data}
We applied TPC to find the CFC from datasets in the public benchmarking platform - \emph{CauseMe} \citep{bussmann2021neural}, and compared with benchmarked approaches in the platform in their performance to recover causal interactions present in the datasets. We used the \emph{River Runoff} (real data) and \emph{Logistic Map} (synthetic data) benchmarking datasets (See Appendix \ref{data:bm}). We compare the approaches PCMCI-GPDC \citep{runge2019detecting}, selVAR \citep{pmlr-v123-weichwald20a} - which are among the top of the leaderboard for performance on the benchmarking datasets, GC, DPGM (estimated by the PC algorithm), and TPC (Our).

\begin{table}[t]
\centering
\begin{tabular}{lll}\toprule
\multicolumn{3}{c}{\textbf{Combined [True, 1-False] Rates (\%)}}\\\midrule
     \textbf{Algorithm} & \textbf{River-Runoff (Real)} & \textbf{Logistic Map  (Synthetic)}\\\midrule
    GC  & ~~~~~37 [45, 92] & ~~~~79 [86, 93]\\
    PCMCI & ~~~~~45 [\underline{100}, 45] & ~~~~86 [89, 97]\\
    selVAR & ~~~~~54 [91, 63] & ~~~~\textbf{87} [88, \underline{99}]\\
    DPGM & ~~~~~60 [64, \underline{96}] & ~~~~19 [27, 92]\\
    TPC (Our) &\textbf{72(+12\%)} [\underline{100}, 72] & 84(-3\%) [\underline{90}, 94]\\\bottomrule
\end{tabular}
\vspace{2mm}
\caption{Comparison of CFC inference by GC, DPGM, PCMCI-GPDC and selVAR, and TPC on benchmarking datasets. For each dataset, each method's Combined Score, True Positive Rate, and 1-False Positive Rate are reported (Higher value is better).}\label{tab:causalsummarybm}
\end{table}

As previously, we measured the performance of the algorithms using 1 - False Positive Rate (IFPR), True Positive Rate (TPR) and Combined Score given by Youden's Index (CS = TPR - FPR) (See Table \ref{tab:causalsummarybm}). The River Runoff dataset is comprised of contemporaneous interactions and is expected to demonstrate the performance of the methods in an empirical setting. The Logistic Map dataset is synthetic and excludes contemporaneous interactions and shows the performance of the methods when the ground truth connectivities are specifically controlled for. 

In terms of CS, TPC has recorded the best performance with a score of $72\%$, followed by DPGM, selVAR, PCMCI-GPDC and GC at $60\%,54\%,45\%, 37\%$ respectively. TPC exceeds the second best approach \textbf{by 12\%}. In terms of TPR, TPC and PCMCI have the highest scores at $100\%$, followed by selVAR, DPGM and GC at $91\%,64\%, 45\%$ respectively. In terms of IFPR, DPGM has the best performance with a score of $96\%$ closely followed by GC at $92\%$, TPC at $72\%$, and selVAR and PCMCI with $63\%, 45\%$. DPGM and GC turn out to have higher IFPR than TPC because they detect fewer false edges, but that is achieved by their detection of fewer edges altogether including fewer true edges leading to a low TPR. In contrast, TPC has greater sensitivity in detecting edges, whose benefit is that TPC detects all the true edges correctly leading to $100\%$ TPR. In terms of both TPR and IFPR, TPC maintains a better trade-off, and thereby a better CS compared to other methods. 

For the logistic map dataset, in terms of CS, selVAR, PCMCI-GPDC and TPC have scores of $87\%, 86\%, 84\%$ respectively, followed by GC with $79\%$ and lastly DPGM with $19\%$. In terms of TPR, TPC has the highest score of $90\%$, followed by PCMCI, selVAR, GC and TPC with TPR of $89\%, 88\%, 86\%, 79\%$ respectively, with DPGM having comparatively lowest TPR of $27\%$. In terms of IFPR, all the approaches have a score of at least $90\%$. Thereby, TPC achieves a high CS of $84\%$, short of $3\%$ from the best CS by selVAR of $87\%$.

The results indicate that in the real benchmark dataset of River-Runoff, TPC outperforms all methods by a substantial gap, whereas, in the synthetic benchmark dataset of Logistic Map, TPC has Combined Score of $84\%$, being in the top group of $84-87\%$ CS performance. While selVAR and PCMCI achieve a CS of $87\%$ and $86\%$ respectively in the synthetic dataset, they achieve a low CS of $54\%$ and $45\%$ in the real dataset. Since the synthetic dataset is generated by a model controlling coupling strength between variables, low noise and devoid of contemporaneous interactions, thereby most of the methods including TPC perform fairly well in the range of $80\%$. In contrast, in the real dataset, the coupling between variables as well as noise are not controlled and contemporaneous interactions are expected to be present as the sampling resolution is greater than the time taken for interactions between the variables. Thereby, the real dataset provides a challenge for the methods where TPC outperforms other approaches with a CS and shows significant improvement in performance than the other methods. Presence of TPC in top group of performance for both benchmarks indicates the generality and applicability of TPC to various scenarios.

\section{Application to Neurobiological Data}
\begin{figure}[t]
    \centering
    \includegraphics[width=1.2\textwidth]{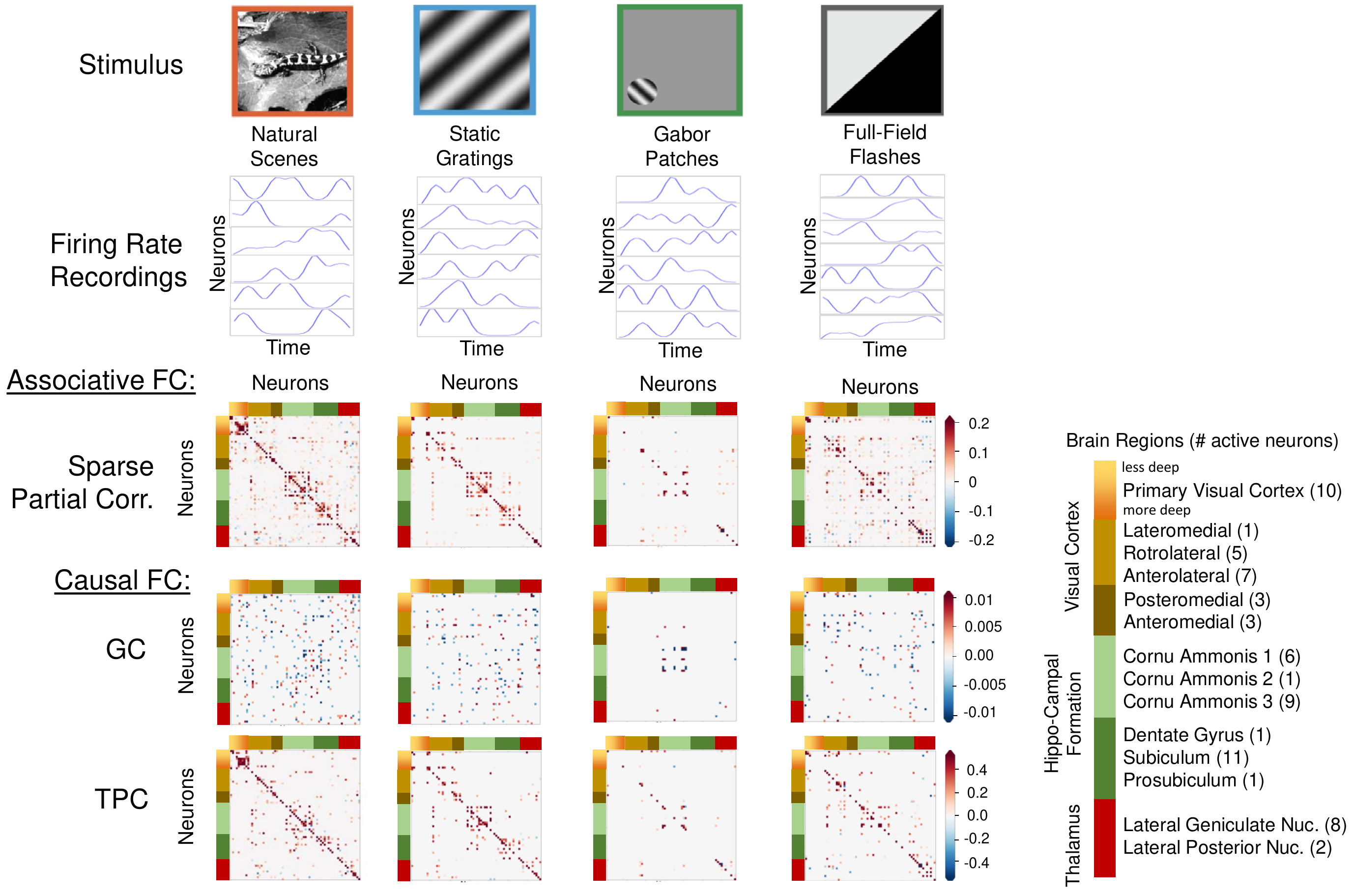}
    \caption{Comparison and demonstration of the FC inferred for a benchmark of mice brain data from the Allen Institute’s Neuropixels dataset, by three methods for FC inference: Associative FC using Sparse Partial Correlation, and Causal FC using GC and TPC. The estimated FC is represented by its adjacency matrix with edge weights, which is symmetric for Associative FC and asymmetric for Causal FC. The mice were subject to different stimuli, among which we selected four stimuli categories with distinct characteristics: Natural Scenes, Static Gratings, Gabor Patches and Full-Field Flashes. The neurons are clustered by the region of brain: Visual Cortex, Hippo-Campal Formation, and Thalamus, which are further divided into sub-regions. In the adjacency matrices, a non-zero entry in $(i,j)$ represents the connection of neuron $i\rightarrow j$.
}
    \label{fig:cfcneuropixels}
\end{figure}
 We proceed and test the methods on neural data consisting of electrophysiological recordings in the Visual Coding Neuropixels dataset of the Allen Brain Observatory ~\citep{de2020large,allenbrainobs}. We compare TPC with Granger Causality (GC) and Sparse Partial Correlation, that are popular methods for obtaining CFC and Associative Functional Connectivity (AFC) from electrophysiological neural recordings. The dataset consists of sorted spike trains and local field potentials recorded simultaneously from up to six cortical visual areas, hippocampus, thalamus, and other adjacent structures of mice, while the mice passively view a stimuli shown to them. The stimuli include static gratings, drifting gratings, natural scenes/images and natural movies, which are shown to the mice with repetitions. The data has been recorded from the neurons with the recently developed technology of Neuropixels which allows real-time recording from hundreds of neurons across the brain simultaneously by inserting multiple probes into the brain~\citep{neuropixels}. Details of the dataset are in Appendix \ref{data:desc}. 

Figure \ref{fig:cfcneuropixels} shows the adjacency matrices for the FC obtained by the methods for one trial in each of the stimuli categories. For each stimuli categories, the AFC constitutes a distinct pattern of connectivity among the neurons. It is expected that the CFC will be a directed subgraph of the AFC and be consistent with the overall patterns present in the AFC \citep{dadgostar2016functional, wang2016efficient}. However, the patterns present in the CFC obtained by GC do not match with the AFC. In contrast, the overall patterns present in the CFC obtained by TPC indeed match with the AFC. On a detailed level, there are differences between TPC-CFC and AFC: TPC results in a directed graph thereby its adjacency matrix is asymmetric while AFC is an undirected graph with symmetric adjacency matrix. Furthermore, the CFC obtained by TPC includes self-loops represented by the diagonals of the adjacency matrix and results in a sparse matrix devoid of noise since the connections passed conditional independence tests and bootstrap stability thresholds. In the CFC obtained by TPC (see Figure \ref{fig:cfcneuropixels}), a greater extent of connectivity within the active neurons in Primary Visual Cortex is evoked by natural scenes, in Posteromedial and Anteromedial Visual Cortex by static gratings, in Anterolateral Visual Cortex and Thalamus by full-field flashes, compared to other stimuli. All four stimuli exhibit distinct patterns of connectivity in the Cornu Ammonis regions of the Hippo-Campal Formation. Natural Scenes and Static gratings evoke more prominent connectivity within the Subiculum compared to other stimuli.

\subsection{Graphical Comparison of Estimated CFC over Stimuli} To study the differences in functional connectivity between the stimuli categories, we investigate the topological patterns in the CFC estimated above. The topological patterns can be summarized by graph theoretic measures \citep{sporns2004organization,van2008small,ueda2018brain}, as follows. The graph measures were computed using the \emph{Networkx} Python library \citep{hagberg2008exploring}, over different trials of each stimuli.

\begin{itemize}
    \item Betweenness centrality: the fraction of all shortest paths that pass through a node, averaged over nodes, indicating the average effect of individual nodes on information flow among the remaining network's nodes.
    \item Transitivity: the fraction of all possible triangles present in the graph, indicating prevalence of clustered connectivity.
    \item Assortativity: measures the similarity of connections in the graph with respect to the node degree.
    \item Clustering Coefficient: the average of all clustering coefficients in the network, reflecting the tightness of connections between nodes.
    \item Global Efficiency: average inverse shortest path length, reflecting node's ability to propagate information with other nodes in the graph.
    \item Local Efficiency: measures the global efficiency for the neighborhood of a node, averaged over nodes, indicating efficiency of transmitting information by nodes with their neighborhood in the graph.
\end{itemize}

\begin{figure}[t!]
    \centering
    \includegraphics[width= \textwidth]{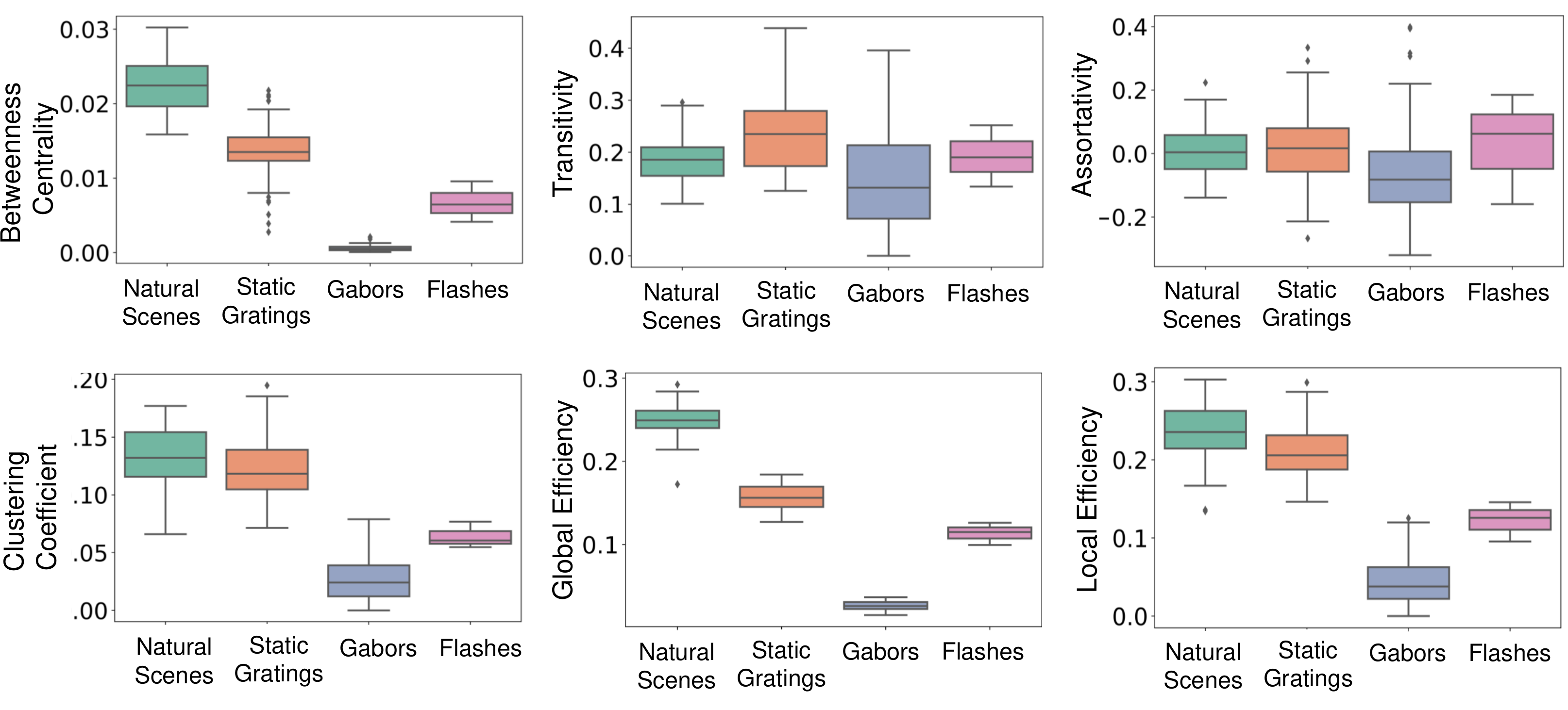}
    \caption{This figure compares the distribution of graph measures of CFC obtained by TPC over different stimuli: natural scenes, static gratings, gabor patches and flashes. The distribution for each graph measure and stimuli is shown by a boxplot.}
    \label{fig:fig7}
\end{figure}

The results of the graph measures for different stimuli are summarized by boxplots. The boxplot of a graph measure (e.g. betweenness centrality) for a stimulus (e.g. natural scenes) shows the distribution of the values of the graph measure over trials for that stimulus, with the top and bottom of the box indicating the upper and lower quartiles of the distribution, the middle of the box indicating the median, while the whiskers extend to show the rest of the distribution excluding outliers, which are marked by points (See Figure \ref{fig:fig7}).

In Figure \ref{fig:fig7}, in terms of \emph{betweenness centrality}, natural scenes have higher score compared to other stimuli, followed by static gratings, then flashes and lastly gabor patches. This shows that during natural scenes, the active neurons have a more remarkable effect on the neural information flow, compared to other stimuli, while gabor patches have the least remarkable effect. In terms of \emph{transitivity}, the scores between the stimuli are close, though static gratings have a relatively higher transitivity, followed by natural scenes, flashes and gabors. This indicates that static gratings evoked a relatively higher prevalence of clustered connectivity, followed by natural scenes, flashes and lastly gabor patches. In terms of \emph{assortativity}, the stimuli did not evoke a comparatively distinct score as well. In terms of \emph{clustering coefficient}, natural scenes and static gratings have higher scores compared to flashes and gabors. This shows that, natural scenes and static gratings have the most tightness of connections between nodes in the graph, flashes have a comparatively lower tightness of connections, while gabor patches have the least score. In terms of \emph{global efficiency}, natural scenes have the highest score, while, static gratings have comparatively lower score, followed by flashes and gabor patches. This shows that natural scenes evoked comparatively highest efficiency of information propagation in the CFC globally, followed by static gratings, flashes and gabors. In terms of \emph{local efficiency}, the trend across stimuli is similar to global efficiency, while natural scenes and static gratings evoked a more similar local efficiency compared to global efficiency. This shows that the efficiency in information propagation in local neighborhoods of neurons is higher for natural scenes and static gratings to a similar extent, but with higher efficiency compared to flashes and gabor patches.

\section{Discussion}

In this paper, we propose a novel methodology, the TPC Algorithm, for finding causal functional connectivity between neurons from neural time series using Directed Probabilistic Graphical Models (DPGM). In particular, we extend the applicability of DPGM to CFC inference from time series by unrolling and implementing the Directed Markov Property (DMP) to obtain the unrolled DAG reflecting causal spatial and temporal interactions. We then roll the DAG back to obtain the CFC graph. The methodology exhibits interpretability of causal interactions over time between neural entities. It also incorporates time delays in interactions between neurons as well as the presence of feedback-loops. The model and the approach are non-parametric, meaning that it does not require the specification of a parametric dynamical equation for neural activity. We show that if the neural activity obeys an arbitrary dynamical process, the Rolled CFC-DPGM is consistent with respect to the causal relationships implied by the dynamical process. We determine that the Rolled CFC-DPGM is predictive of counterfactual queries such as ablation or modulation. We show that the answers can be provided by using simple causal reasoning with the edges of the rolled CFC-DPGM. We demonstrate the utilization of the methodology in obtaining CFC from simulations and compare the performance of TPC with other methods such as Granger Causality (GC) and common DPGM. Furthermore, we apply the methods to benchmarks of time-series causal inference and neurobiological dataset from mice brain presented with various visual stimuli. The results provide insights into the CFC between neurons in the mouse brain in a variety of stimuli scenario. We also compare the topological patterns in the estimated CFC between different stimuli using graph-theoretic measures.

\begin{table}[t!]
\caption{Comparative summary of different approaches for causal modeling.}
\label{tab:causalsummary}
\begin{tabular}{p{2.5cm} p{2.4cm} p{2.4cm} p{2.4cm} p{2.8cm} p{2.8cm}}\toprule
     & \textbf{GC} & \textbf{DCM} & \textbf{DPGM} & \textbf{TPC} \\\midrule
    Form of Causality & Non-zero parameters in VAR model & Coupling parameters in biological model & Directed Markov Graph & Directed Markov Graph over time-delayed variables \\\midrule
    
    Inclusion of temporal relationships & \textcolor{ForestGreen}{\textbf{Yes}} & \textcolor{ForestGreen}{\textbf{Yes}} & \textcolor{Maroon}{\textbf{No}}, formulation for static variables & \textcolor{ForestGreen}{\textbf{Yes}}, adapts DPGM for inter-temporal relationships\\\midrule
    Inclusion of contemporaneous relationships & \textcolor{Maroon}{\textbf{No}} & \textcolor{ForestGreen}{\textbf{Yes}}, by a differential equation  & \textcolor{Maroon}{\textbf{No}} & \textcolor{ForestGreen}{\textbf{Yes}}, if $(i,t)\rightarrow (j,t)$ then $i\rightarrow j$.\\\midrule
    Generalizable Statistical Model & \textcolor{ForestGreen}{\textbf{Yes}} & \textcolor{Maroon}{\textbf{No}} & \textcolor{ForestGreen}{\textbf{Yes}} & \textcolor{ForestGreen}{\textbf{Yes}}\\\midrule
    Non-parametric Model & \textcolor{ForestGreen}{\textbf{Yes}}, parametric and non-parametric approaches exist. & \textcolor{Maroon}{\textbf{No}}, biologically mechanistic non-linear model. & \textcolor{ForestGreen}{\textbf{Yes}}, equivalent to an arbitrary functional relationship between nodes. & \textcolor{ForestGreen}{\textbf{Yes}}, equivalent to an arbitrary functional relationship between neural activity at different times.\\\midrule
    Supports CFC Inference & \textcolor{ForestGreen}{\textbf{Yes}} & \textcolor{Maroon}{\textbf{No}}, suitable for comparing model hypotheses & \textcolor{ForestGreen}{\textbf{Yes}} & \textcolor{ForestGreen}{\textbf{Yes}}\\\midrule
    Cycles (including self-loops) Occuring In The Model & \textcolor{ForestGreen}{\textbf{Yes}} for VAR model (neuron $i\rightarrow i$ when $A_{ii}(k)\neq 0$ for some $k$). & \textcolor{ForestGreen}{\textbf{Yes}} ($i\rightarrow i$ when $\theta_{ii}\neq 0$) & \textcolor{Maroon}{\textbf{No}}, it is a DAG & \textcolor{ForestGreen}{\textbf{Yes}} ($i\rightarrow i$ when $(i,t)\rightarrow (i,t')$ for some $t<t'$)\\\midrule
    Incorporation of Interventional and Counterfactual queries & \textcolor{Maroon}{\textbf{No}} & \textcolor{Maroon}{\textbf{No}} & \textcolor{ForestGreen}{\textbf{Yes}} but for static variables. & \textcolor{ForestGreen}{\textbf{Yes}}, adapts for temporal scenario, can predict the consequence on CFC of counterfactual intervention to neural activity.\\
    \bottomrule
\end{tabular}
\end{table}

The distinctive and useful aspect of TPC is that it takes into consideration neural interactions over time. In neuroscience literature, causality is typically referred to as ``a cause of an observed neural event (the ‘effect’) as a preceding neural event whose occurrence is necessary to observe the effect"~\citep{reid2019advancing}. The approach of unrolling causal graphs over time and considering time-delays in TPC incorporates this definition by essentially finding whether the previous time values of the neurons impact the present value of a particular neuron by the obtained directed graph. By virtue of the Directed Markov Property, TPC incorporates causality of neural interactions in a non-parameteric, model-free manner and incorporates interventional properties. The bootstrapping step filters spurious connections by repeating the inference of CFC over several subsampled blocks of the time series and discarding those connections that are absent in multiple repetitions. Pruning further selects edges by discarding edges with exceedingly low edge weight. In contrast, parametric approaches, such as GC, do not satisfy the Directed Markov Property, thereby not incorporating interventional properties. Furthermore, in comparative studies with simulated, benchmark and real neurobiological datasets, we found the performance of TPC to be better compared to other approaches. We conclude by summarizing, in Table \ref{tab:causalsummary}, the differences and benefits of TPC in a comparison with other approaches including variants of the PC algorithm (DPGM) which satisfy DMP in a static data setting, and other approaches that do not obey the DMP, such as Granger Causality (GC) and Dynamic Causal Model (DCM), outlining their strengths and weaknesses with respect to several criteria of causality in functional connectomics. Indeed, capturing as many causal criteria is fundamental to any approach from statistical and application points of view. 

Our exposition of properties of each approach and the comparative study show that each of the methods address different aspects of modeling causality of neural interaction and mapping them in the form of a graph \citep{biswas2021statistical}. The comparative table demonstrates that with respect to the model that each approach is assuming, GC requires a linear model in its common use, though recent non-linear and non-parametric extensions, have been applied. DCM requires a strict well defined mechanistic biological model and thus can only compare different models based on evidence from data. In comparison, DPGM and TPC have the advantage of not requiring modeling of the neural dynamics using a parametric equation or assumption of a linear model. While DPGM is developed for static variables, and as such cannot address temporal and contemporaneous relationships and must obey DAG architechture, TPC is suited for time series setting and extends DPGM for spatiotemporal data. TPC obtains the CFC that follows Directed Markov Property extended to include inter-temporal relationships in a time-series setting such that parent-child relations between neurons are equivalent to arbitrary functional relationships between their neural activity over time. In terms of incorporating contemporaneous interactions arising when causal interactions happen faster than sampling rate, while GC and DPGM do not, TPC is able to incorporate contemporaneous interactions by design. In terms of incorporating self-loops in neural activity, while DPGM typically produces a DAG, TPC incorporates self-loops in neural activity. In regards to guarantee of causality, GC can provide useful insights into a system's dynamical interactions in different conditions, however its causal interpretation is not guaranteed as it focuses on the predictability of future based on past observations of variables. DCM uses the parameters for coupling between hidden neural states in competing biological models to indicate CFC, however it compares hypothetical models based on evidence from data which relevance to causality is not guaranteed (Friston et al., 2003). In summary, TPC extends DPGM to the time-series setting and provides a probabilistic foundation for causality which is predictive of the consequence of possible intervention like neuron ablation and neuromodulation.

While TPC provides a powerful causal framework for time series, its current version relies on the PC algorithm which in turn assumes \emph{causal sufficiency}, that is, all the causes of the input variables are present within the input variables. In the presence of latent confounders, the PC can lead to spurious edges. These are partly remediated by the TPC Algorithm by discarding spurious edges in the Bootstrap and Pruning steps, but some of these effect may remain. To address this, an alternative strategy that could be considered is replacing the PC algorithm with the FCI algorithm in Step 3 of TPC (\ref{algo:TPC}) since the FCI algorithm exhibits statistical consistency, that converges in probability to the true causal relationships, given i.i.d. samples in the presence of latent confounders \citep{spirtes1999algorithm}. However, the FCI algorithm identifies indirect connections only and not direct connections, thereby would result in a CFC with indirect causal connections. 

Under the assumptions of 1) causal sufficiency, 2) faithfulness, and 3) given i.i.d. samples, the PC algorithm's estimated causal DAG is statistically consistent. In light of these assumptions, the step 2 of the TPC algorithm constructs samples with a time-delay of $2(\tau+1)$ between samples from a bootstrap window, for max time-delay of interaction $\tau$, which are used as an input to the PC algorithm in the next step. The time-delay between samples of $2(\tau+1)$ reduces between-sample dependence. And, the samples being constructed from a short bootstrap window instead of the entire time series aids to make their distribution fairly identical. Yet, specifications of $\tau$ of lower value could lead to between-sample dependence and rapid perturbations to the time series in a bootstrap window can lead to distribution changes between samples in a bootstrap window, leading to potential reduction in efficacy of edge detection by the PC algorithm. To improve overall efficacy in such scenarios and curb spurious edge detections, the bootstrap step of the TPC algorithm outputs many CFCs over random time windows and preserves only stable edges over the set of CFCs. In the pruning step, the edges in the CFC are pruned if having an exceedingly low connectivity weight.

In conclusion, TPC provides a probabilistic and interpretable formulation for CFC modeling and inference in the context of neural time series. We have established the statistical properties of the model as well as demonstrated its performance in estimation of CFC. We have demonstrated TPC application in continuous time series datasets, however TPC is similarly applicable to discrete time series datasets by simply using a statistical conditional independence test for discrete data in the algorithm.  This can be especially relevant for count datasets such as spiking neuron datasets.

\bibliographystyle{unsrt}
%\bibliography{main}

\begin{thebibliography}{10}

\bibitem{reid2012functional}
R~Clay Reid.
\newblock From functional architecture to functional connectomics.
\newblock {\em Neuron}, 75(2):209--217, 2012.

\bibitem{reid2019advancing}
Andrew~T Reid, Drew~B Headley, Ravi~D Mill, Ruben Sanchez-Romero, Lucina~Q
  Uddin, Daniele Marinazzo, Daniel~J Lurie, Pedro~A Vald{\'e}s-Sosa,
  Stephen~Jos{\'e} Hanson, Bharat~B Biswal, et~al.
\newblock Advancing functional connectivity research from association to
  causation.
\newblock {\em Nature neuroscience}, 1(10), 2019.

\bibitem{cassidy2021functional}
Jessica~M Cassidy, Jasper~I Mark, and Steven~C Cramer.
\newblock Functional connectivity drives stroke recovery: shifting the paradigm
  from correlation to causation.
\newblock {\em Brain}, 2021.

\bibitem{sanchez2021combining}
Ruben Sanchez-Romero and Michael~W Cole.
\newblock Combining multiple functional connectivity methods to improve causal
  inferences.
\newblock {\em Journal of cognitive neuroscience}, 33(2):180--194, 2021.

\bibitem{finn2015functional}
Emily~S Finn, Xilin Shen, Dustin Scheinost, Monica~D Rosenberg, Jessica Huang,
  Marvin~M Chun, Xenophon Papademetris, and R~Todd Constable.
\newblock Functional connectome fingerprinting: identifying individuals using
  patterns of brain connectivity.
\newblock {\em Nature neuroscience}, 18(11):1664--1671, 2015.

\bibitem{biswas2021statistical}
Rahul Biswas and Eli Shlizerman.
\newblock Statistical perspective on functional and causal neural connectomics:
  A comparative study.
\newblock {\em Frontiers in Systems Neuroscience}, 16, 2022.

\bibitem{sharaev2016effective}
Maksim~G Sharaev, Viktoria~V Zavyalova, Vadim~L Ushakov, Sergey~I Kartashov,
  and Boris~M Velichkovsky.
\newblock Effective connectivity within the default mode network: dynamic
  causal modeling of resting-state fmri data.
\newblock {\em Frontiers in human neuroscience}, 10:14, 2016.

\bibitem{lauritzen2001causal}
Steffen~L Lauritzen.
\newblock Causal inference from graphical models.
\newblock {\em Complex stochastic systems}, pages 63--107, 2001.

\bibitem{maathuis2018handbook}
Marloes Maathuis, Mathias Drton, Steffen Lauritzen, and Martin Wainwright.
\newblock {\em Handbook of graphical models}.
\newblock CRC Press, 2018.

\bibitem{gomez2020functional}
Ana Mar{\'\i}a~Estrada G{\'o}mez, Kamran Paynabar, and Massimo Pacella.
\newblock Functional directed graphical models and applications in root-cause
  analysis and diagnosis.
\newblock {\em Journal of Quality Technology}, pages 1--17, 2020.

\bibitem{ahelegbey2016econometrics}
Daniel~Felix Ahelegbey.
\newblock The econometrics of bayesian graphical models: a review with
  financial application.
\newblock {\em Journal of Network Theory in Finance}, 2(2):1--33, 2016.

\bibitem{ebert2012causal}
Imme Ebert-Uphoff and Yi~Deng.
\newblock Causal discovery for climate research using graphical models.
\newblock {\em Journal of Climate}, 25(17):5648--5665, 2012.

\bibitem{kalisch2010understanding}
Markus Kalisch, Bernd~AG Fellinghauer, Eva Grill, Marloes~H Maathuis, Ulrich
  Mansmann, Peter B{\"u}hlmann, and Gerold Stucki.
\newblock Understanding human functioning using graphical models.
\newblock {\em BMC Medical Research Methodology}, 10(1):1--10, 2010.

\bibitem{deng2005structural}
Ke~Deng, Delin Liu, Shan Gao, and Zhi Geng.
\newblock Structural learning of graphical models and its applications to
  traditional chinese medicine.
\newblock In {\em International Conference on Fuzzy Systems and Knowledge
  Discovery}, pages 362--367. Springer, 2005.

\bibitem{haigh2004causality}
Michael~S Haigh and David~A Bessler.
\newblock Causality and price discovery: An application of directed acyclic
  graphs.
\newblock {\em The Journal of Business}, 77(4):1099--1121, 2004.

\bibitem{wang2017potential}
Huange Wang, Fred~A van Eeuwijk, and Johannes Jansen.
\newblock The potential of probabilistic graphical models in linkage map
  construction.
\newblock {\em Theoretical and Applied Genetics}, 130(2):433--444, 2017.

\bibitem{sinoquet2014probabilistic}
Christine Sinoquet.
\newblock {\em Probabilistic graphical models for genetics, genomics, and
  postgenomics}.
\newblock OUP Oxford, 2014.

\bibitem{mourad2012probabilistic}
Rapha{\"e}l Mourad, Christine Sinoquet, and Philippe Leray.
\newblock Probabilistic graphical models for genetic association studies.
\newblock {\em Briefings in bioinformatics}, 13(1):20--33, 2012.

\bibitem{wang2005new}
Junbai Wang, Leo Wang-Kit Cheung, and Jan Delabie.
\newblock New probabilistic graphical models for genetic regulatory networks
  studies.
\newblock {\em Journal of biomedical informatics}, 38(6):443--455, 2005.

\bibitem{liu2018functional}
Hexuan Liu, Jimin Kim, and Eli Shlizerman.
\newblock Functional connectomics from neural dynamics: probabilistic graphical
  models for neuronal network of caenorhabditis elegans.
\newblock {\em Philosophical Transactions of the Royal Society B: Biological
  Sciences}, 373(1758):20170377, 2018.

\bibitem{friedman2004inferring}
Nir Friedman.
\newblock Inferring cellular networks using probabilistic graphical models.
\newblock {\em Science}, 303(5659):799--805, 2004.

\bibitem{spirtes2000causation}
Peter Spirtes, Clark~N Glymour, Richard Scheines, and David Heckerman.
\newblock {\em Causation, prediction, and search}.
\newblock MIT press, 2000.

\bibitem{pearl2009causality}
Judea Pearl.
\newblock {\em Causality}.
\newblock Cambridge university press, 2009.

\bibitem{fiete2010spike}
Ila~R Fiete, Walter Senn, Claude~ZH Wang, and Richard~HR Hahnloser.
\newblock Spike-time-dependent plasticity and heterosynaptic competition
  organize networks to produce long scale-free sequences of neural activity.
\newblock {\em Neuron}, 65(4):563--576, 2010.

\bibitem{arbabshirani2019autoconnectivity}
Mohammad~R Arbabshirani, Adrian Preda, Jatin~G Vaidya, Steven~G Potkin, Godfrey
  Pearlson, James Voyvodic, Daniel Mathalon, Theo van Erp, Andrew Michael,
  Kent~A Kiehl, et~al.
\newblock Autoconnectivity: A new perspective on human brain function.
\newblock {\em Journal of neuroscience methods}, 323:68--76, 2019.

\bibitem{borisyuk1999oscillatory}
Roman Borisyuk and Frank Hoppensteadt.
\newblock Oscillatory models of the hippocampus: a study of spatio-temporal
  patterns of neural activity.
\newblock {\em Biological cybernetics}, 81(4):359--371, 1999.

\bibitem{jutras2010synchronous}
Michael~J Jutras and Elizabeth~A Buffalo.
\newblock Synchronous neural activity and memory formation.
\newblock {\em Current opinion in neurobiology}, 20(2):150--155, 2010.

\bibitem{richardson1996automated}
Thomas~S Richardson, Peter Spirtes, et~al.
\newblock {\em Automated discovery of linear feedback models}.
\newblock Carnegie Mellon [Department of Philosophy], 1996.

\bibitem{bollen1989structural}
Kenneth~A Bollen.
\newblock Structural equations with latent variables wiley.
\newblock {\em New York}, 1989.

\bibitem{drton2017structure}
Mathias Drton and Marloes~H Maathuis.
\newblock Structure learning in graphical modeling.
\newblock {\em Annual Review of Statistics and Its Application}, 4:365--393,
  2017.

\bibitem{kalisch2007estimating}
Markus Kalisch and Peter B{\"u}hlmann.
\newblock Estimating high-dimensional directed acyclic graphs with the
  pc-algorithm.
\newblock {\em Journal of Machine Learning Research}, 8(Mar):613--636, 2007.

\bibitem{tillman2009nonlinear}
Robert~E Tillman, Arthur Gretton, and Peter Spirtes.
\newblock Nonlinear directed acyclic structure learning with weakly additive
  noise models.
\newblock In {\em NIPS}, pages 1847--1855. Citeseer, 2009.

\bibitem{spirtes1999algorithm}
Peter Spirtes, Christopher Meek, and Thomas Richardson.
\newblock An algorithm for causal inference in the presence of latent variables
  and selection bias.
\newblock {\em Computation, causation, and discovery}, 21:1--252, 1999.

\bibitem{zhang2012strong}
Jiji Zhang and Peter~L Spirtes.
\newblock Strong faithfulness and uniform consistency in causal inference.
\newblock {\em arXiv preprint arXiv:1212.2506}, 2012.

\bibitem{swanson1997impulse}
Norman~R Swanson and Clive~WJ Granger.
\newblock Impulse response functions based on a causal approach to residual
  orthogonalization in vector autoregressions.
\newblock {\em Journal of the American Statistical Association},
  92(437):357--367, 1997.

\bibitem{runge2019inferring}
Jakob Runge, Sebastian Bathiany, Erik Bollt, Gustau Camps-Valls, Dim Coumou,
  Ethan Deyle, Clark Glymour, Marlene Kretschmer, Miguel~D Mahecha, Jordi
  Mu{\~n}oz-Mar{\'\i}, et~al.
\newblock Inferring causation from time series in earth system sciences.
\newblock {\em Nature communications}, 10(1):1--13, 2019.

\bibitem{shinomoto2010estimating}
Shigeru Shinomoto.
\newblock Estimating the firing rate.
\newblock In {\em Analysis of Parallel Spike Trains}, pages 21--35. Springer,
  2010.

\bibitem{stevenson2008inferring}
Ian~H Stevenson, James~M Rebesco, Lee~E Miller, and Konrad~P K{\"o}rding.
\newblock Inferring functional connections between neurons.
\newblock {\em Current opinion in neurobiology}, 18(6):582--588, 2008.

\bibitem{stephan2007comparing}
Klaas~Enno Stephan, Nikolaus Weiskopf, Peter~M Drysdale, Peter~A Robinson, and
  Karl~J Friston.
\newblock Comparing hemodynamic models with dcm.
\newblock {\em Neuroimage}, 38(3):387--401, 2007.

\bibitem{mullins2016unifying}
Caitlin Mullins, Gord Fishell, and Richard~W Tsien.
\newblock Unifying views of autism spectrum disorders: a consideration of
  autoregulatory feedback loops.
\newblock {\em Neuron}, 89(6):1131--1156, 2016.

\bibitem{sheffield2016cognition}
Julia~M Sheffield and Deanna~M Barch.
\newblock Cognition and resting-state functional connectivity in schizophrenia.
\newblock {\em Neuroscience \& Biobehavioral Reviews}, 61:108--120, 2016.

\bibitem{pearl2009causal}
Judea Pearl et~al.
\newblock Causal inference in statistics: An overview.
\newblock {\em Statistics surveys}, 3:96--146, 2009.

\bibitem{maathuis2009estimating}
Marloes~H Maathuis, Markus Kalisch, Peter B{\"u}hlmann, et~al.
\newblock Estimating high-dimensional intervention effects from observational
  data.
\newblock {\em The Annals of Statistics}, 37(6A):3133--3164, 2009.

\bibitem{huang2012pearl}
Yimin Huang and Marco Valtorta.
\newblock Pearl's calculus of intervention is complete.
\newblock {\em arXiv preprint arXiv:1206.6831}, 2012.

\bibitem{vsimundic2009measures}
Ana-Maria {\v{S}}imundi{\'c}.
\newblock Measures of diagnostic accuracy: basic definitions.
\newblock {\em Ejifcc}, 19(4):203, 2009.

\bibitem{hilden1996regret}
J{\o}rgen Hilden and Paul Glasziou.
\newblock Regret graphs, diagnostic uncertainty and youden's index.
\newblock {\em Statistics in medicine}, 15(10):969--986, 1996.

\bibitem{bussmann2021neural}
Bart Bussmann, Jannes Nys, and Steven Latr{\'e}.
\newblock Neural additive vector autoregression models for causal discovery in
  time series.
\newblock In {\em International Conference on Discovery Science}, pages
  446--460. Springer, 2021.

\bibitem{runge2019detecting}
Jakob Runge, Peer Nowack, Marlene Kretschmer, Seth Flaxman, and Dino
  Sejdinovic.
\newblock Detecting and quantifying causal associations in large nonlinear time
  series datasets.
\newblock {\em Science Advances}, 5(11):eaau4996, 2019.

\bibitem{pmlr-v123-weichwald20a}
Sebastian Weichwald, Martin~E. Jakobsen, Phillip~B. Mogensen, Lasse Petersen,
  Nikolaj Thams, and Gherardo Varando.
\newblock Causal structure learning from time series: Large regression
  coefficients may predict causal links better in practice than small p-values.
\newblock In Hugo~Jair Escalante and Raia Hadsell, editors, {\em Proceedings of
  the NeurIPS 2019 Competition and Demonstration Track}, volume 123 of {\em
  Proceedings of Machine Learning Research}, pages 27--36. PMLR, 08--14 Dec
  2020.

\bibitem{de2020large}
Saskia~EJ de~Vries, Jerome~A Lecoq, Michael~A Buice, Peter~A Groblewski,
  Gabriel~K Ocker, Michael Oliver, David Feng, Nicholas Cain, Peter
  Ledochowitsch, Daniel Millman, et~al.
\newblock A large-scale standardized physiological survey reveals functional
  organization of the mouse visual cortex.
\newblock {\em Nature Neuroscience}, 23(1):138--151, 2020.

\bibitem{allenbrainobs}
Allen-Brain-Observatory.
\newblock Allen institute for brain science.
\newblock {\em Available from:
  https://portal.brain-map.org/explore/circuits/visual-coding-neuropixels},
  October, 2019.

\bibitem{neuropixels}
James~J Jun, Nicholas~A Steinmetz, Joshua~H Siegle, Daniel~J Denman, Marius
  Bauza, Brian Barbarits, Albert~K Lee, Costas~A Anastassiou, Alexandru Andrei,
  {\c{C}}a{\u{g}}atay Ayd{\i}n, et~al.
\newblock Fully integrated silicon probes for high-density recording of neural
  activity.
\newblock {\em Nature}, 551(7679):232--236, 2017.

\bibitem{dadgostar2016functional}
Mehrdad Dadgostar, Seyed~Kamaledin Setarehdan, Sohrab Shahzadi, and Ata Akin.
\newblock Functional connectivity of the pfc via partial correlation.
\newblock {\em Optik}, 127(11):4748--4754, 2016.

\bibitem{wang2016efficient}
Yikai Wang, Jian Kang, Phebe~B Kemmer, and Ying Guo.
\newblock An efficient and reliable statistical method for estimating
  functional connectivity in large scale brain networks using partial
  correlation.
\newblock {\em Frontiers in neuroscience}, 10:123, 2016.

\bibitem{sporns2004organization}
Olaf Sporns, Dante~R Chialvo, Marcus Kaiser, and Claus~C Hilgetag.
\newblock Organization, development and function of complex brain networks.
\newblock {\em Trends in cognitive sciences}, 8(9):418--425, 2004.

\bibitem{van2008small}
Martijn~P van~den Heuvel, Cornelis~J Stam, Maria Boersma, and HE~Hulshoff Pol.
\newblock Small-world and scale-free organization of voxel-based resting-state
  functional connectivity in the human brain.
\newblock {\em Neuroimage}, 43(3):528--539, 2008.

\bibitem{ueda2018brain}
Issei Ueda, Shingo Kakeda, Keita Watanabe, Koichiro Sugimoto, Natsuki Igata,
  Junji Moriya, Kazuhiro Takemoto, Asuka Katsuki, Reiji Yoshimura, Osamu Abe,
  et~al.
\newblock Brain structural connectivity and neuroticism in healthy adults.
\newblock {\em Scientific reports}, 8(1):1--8, 2018.

\bibitem{hagberg2008exploring}
Aric Hagberg, Pieter Swart, and Daniel S~Chult.
\newblock Exploring network structure, dynamics, and function using networkx.
\newblock Technical report, Los Alamos National Lab.(LANL), Los Alamos, NM
  (United States), 2008.

\bibitem{rokem2009nitime}
Ariel Rokem, M~Trumpis, and F~Perez.
\newblock Nitime: time-series analysis for neuroimaging data.
\newblock In {\em Proceedings of the 8th Python in Science Conference}, pages
  68--75, 2009.

\bibitem{schmidt2016multivariate}
Christoph Schmidt, Britta Pester, Nicole Schmid-Hertel, Herbert Witte, Axel
  Wism{\"u}ller, and Lutz Leistritz.
\newblock A multivariate granger causality concept towards full brain
  functional connectivity.
\newblock {\em PloS one}, 11(4):e0153105, 2016.

\end{thebibliography}

\appendix
\section{Proof of Theorem 1} \label{proof:thm1}

Let $\bm{V}=\{(v,t):v\in V, t\in T\}$, $\bm{E}=\{(u_{v,i},t_{v,i})\rightarrow (v,t):1\leq i \leq K, v\in V, t\in [0,T]\}$, and $\bm{G}=(\bm{V},\bm{E})$. 
Rewriting Eq. (\ref{eq:thmfirst}) we get,

\[
\bm{X}(v,t) = g_{v,t} (\{\bm{X}(u,k):(u,k)\in pa_{\bm{G}}((v,t))\}, \epsilon_v(t))
\]

By Theorem 1.4.1 in \citep{pearl2009causality}, the above implies that $\bm{X}$ satisfies the DPM with respect to $\bm{G}$.

Therefore by Definition \ref{def:causalfc_td}, the rolled CFC-DPGM, $F_{\tau}$, has nodes $V$ and edges given by $u_{v,i}\rightarrow v$, since $(u_{v,i},t_{v,i})\rightarrow (v,t) \in \bm{E}$, for $v\in V$ and $1\leq i\leq K$. That is, $pa_{F_{\tau}}(v) = \{u_{v,1},\ldots, u_{v,K}\}$.
\section{Proof of Theorem 2} \label{proof:thm2}
During the experimental/counterfactual intervention, such as controlling the activity of neuron or neuron ablation, the activity $X_{v_1} (t),\ldots, X_{v_k} (t)$ would no longer be a function of the activities of the neurons at preceding time points for $t\in T_I$, while the activity of neurons which are not intervened, would still be a function of activity at preceding time points. That is, $X_v (t) = g_{v,t} (X_{pa_{F_{\tau}}(v)}(t-), \epsilon_{v}(t))$, for $v \notin \{v_1,\ldots,v_k\}$ and $X_v (t) = g_{v_i,t} (\epsilon_{v_i}(t))$ for $v \in \{v_1,\ldots,v_k\}$ for $t\in T_I$, where $g_{v_i,t}(\epsilon_{v_i}(t))$ represents the distribution of neural activity due to the experimental intervention. For example, $g_{v_i,t}$ is identically $0$ for neuron ablation, and can be an oscillating function with a high amplitude and random noise $\epsilon_{v_i}(t)$ for stimulation through external control. The dynamics can be written as \begin{equation}\label{eq:intervention}X_v (t) = g_{v,t} (X_{pa_{F_{\tau}^{I}}(v)}(t-), \epsilon_{v}(t))\end{equation}
for $v \in V, t\in T_I$, where $pa_{F_{\tau}^{I}}(v)=pa_{G}(v)$ for $v\notin \{v_1,\ldots, v_k\}$ and $pa_{F_{\tau}^{I}}(v_1)=\ldots=pa_{F_{\tau}^{I}}(v_k)=\Phi$, where $\Phi$ denotes the null set. 
%That is, $G_I$ is same as $G$ except for the parents of $v_1,\ldots, v_k$. 
In other words, $F_{\tau}^{I}$ has all connections same as $F_{\tau}$ except that all connections directing to the intervened neurons $v_1,\ldots, v_k$ are removed. It follows from Eq. (\ref{eq:intervention}) and Theorem \ref{thm:causal-fc-td} that $F_{\tau}^{I}$ is the causal functional connectivity between the neurons in $V$ during time $T_I$ when $v_1,\ldots,v_k$ are subject to experimental intervention.

For ablations,the edges originating from $v_1,\ldots,v_k$, can be removed since the activity of these neurons would be fixed (at zero) and would be trivial variables that can be excluded from the argument of the function $g_{v,t}$ in Eq. (\ref{eq:intervention}).

\section{Simulation Study Details} \label{simul:details}
We study the following simulation paradigms.
\begin{enumerate}[wide=0pt]
\item \underline{Linear Gaussian Time Series (Figure \ref{fig:simul_eval}a left-column)}. Let $N(0,\eta)$ denote a Normal random variable with mean $0$ and standard deviation $\eta$. We define $X_v(t)$ as a linear Gaussian time series for $v=1,\ldots,4$ whose true CFC has the edges $1\rightarrow 3,2\rightarrow 3, 3\rightarrow 4$. Let $X_v(0)=N(0,\eta)$ for $v=1,\ldots,4$, and for $t=1,2,\ldots,1000$,
    \begin{align*}
    &X_1(t)=1+N(0,\eta),~&&X_2(t)=-1+N(0,\eta),\\
    &X_3(t+1)=2X_1(t)+X_2(t)+N(0,\eta),~&&X_4(t+1)=2X_3(t)+N(0,\eta).
    \end{align*}
    We obtain 25 simulations of the entire time series each for different noise levels $\eta \in \{0.1,0.5,1,1.5,2,2.5,3,3.5\}$.
    \item \underline{Non-linear Non-Gaussian Time Series (Figure \ref{fig:simul_eval}a middle-column)}. Let $U(0,\eta)$ denote a \emph{Uniformly} distributed random variable on the interval $(0,\eta)$. We define $X_v(t)$ as a non-linear non-Gaussian time series for $v=1,\ldots,4$ whose true CFC has the edges $1\rightarrow 3, 2\rightarrow 3, 3\rightarrow 4$. Let $X_v(0)=U(0,\eta)$ for $v=1,\ldots,4$ and for $t=1,2,\ldots,1000$,
    \begin{align*}
    &X_1(t)=U(0,\eta),~&&X_2(t)=U(0,\eta),\\
    &X_3(t+1)=4\sin(X_1(t))+3\cos(X_2(t))+U(0,\eta),~&&X_4(t+1)=2\sin(X_3(t))+U(0,\eta).
    \end{align*}
    We obtain 25 simulations of the entire time series each for different noise levels $\eta \in \{0.1,0.5,1,1.5,2,2.5,3,3.5\}$.

  \item \underline{Continuous Time Recurrent Neural Network (CTRNN) (Figure \ref{fig:simul_eval}a right-column)}. We simulate neural dynamics by Continuous Time Recurrent Neural Networks, Eq. (\ref{ctrnn}). $u_j(t)$ is the instantaneous firing rate at time $t$ for a post-synaptic neuron $j$, $w_{ij}$ is the linear coefficient to pre-synaptic neuron $i$'s input on the post-synaptic neuron $j$, $I_j(t)$ is the input current on neuron $j$ at time $t$, $\tau_j$ is the time constant of the post-synaptic neuron $j$, with $i,j$ being indices for neurons with $m$ being the total number of neurons. Such a model is typically used to simulate neurons as firing rate units,
    \begin{equation}\label{ctrnn}
    \tau_j \frac{du_j(t)}{dt}=-u_j (t) + \sum_{i=1}^m w_{ij} \sigma (u_i (t)) + I_j (t), j=1,\ldots, m.
    \end{equation}
    We consider a motif consisting of $4$ neurons with $w_{13}=w_{23}=w_{34}=10$ and $w_{ij}=0$ otherwise. We also note that in Eq. \ref{ctrnn}, activity of each neuron $u_j(t)$ depends on its own past. Therefore, the true CFC has the edges $1\rightarrow 3,2\rightarrow 3,3\rightarrow 4, 1\rightarrow 1, 2\rightarrow 2, 3\rightarrow 3, 4\rightarrow 4$. The time constant $\tau_i$ is set to 10 msecs for each neuron $i$. We consider $I_i(t)$ to be distributed as independent Gaussian process with the mean of 1 and the standard deviation of $\eta$. The signals are sampled at a time gap of $e \approx 2.72$ msecs for a total duration of $1000$ msecs.     We obtain 25 simulations of the entire time series each for different noise levels $\eta \in \{0.1,0.5,1,1.5,2,2.5,3,3.5\}$.

\end{enumerate}

The GC graph is computed using the \emph{Nitime} Python library, which fits an MVAR model followed by using the \emph{GrangerAnalyzer} to compute the Granger Causality \citep{rokem2009nitime}. The PC algorithm, which requires several samples of a scalar-valued random variable $Y_v$ (measured activity) for neurons $v\in V$, is used to compute DPGM. We define $Y_v$ as a windowed average of recordings over a duration of $50$ msec: $Y_v = X_v, v\in V$, and averaging over different $50$ msec windows with a gap of $50$ msec between consecutive windows yields different $Y_v$ samples. This choice of $Y_v$ performs better than considering $Y_v$ to be neural recordings at time $t$: $Y_v = X_v(t), v\in V$, with different $t$ giving different samples of $Y_v$ in previous work \citep{biswas2021statistical}. The TPC algorithm computes the rolled CFC-DPGM directly from the signals and, we use a maximum time-delay of interaction of $1$ msec.

The choice of thresholds tunes the decision whether a connection exists in the CFC. For DPGM and TPC, increasing the significance level $\alpha$ for conditional independence tests increases the rate of detecting edges, but also increasing the rate of detecting false positives. We consider $\alpha = 0.01, 0.05, 0.1$ for DPGM and TPC. For GC, a likelihood ratio statistic $L_{uv}$ is obtained for testing $A_{uv}(k) = 0$ for $k=1,\ldots,K$. An edge $u\rightarrow v$  is outputted if $L_{uv}$ has a value greater than a threshold. We use a percentile-based threshold, and output an edge $u\rightarrow v$ if $L_{uv}$ is greater than $100(1-\alpha)$ percentile of $L_{ij}$’s over all pairs of neurons $(i,j)$ in the graph \citep{schmidt2016multivariate}. We consider $\alpha = 0.01, 0.05, 0.1$ which corresponds to percentile thresholds of $99\%, 95\%, 90\%$. For the bootstrap procedure in TPC, we consider 50 bootstrap iterations with bootstrap window length of $50$ msec and bootstrap stability cutoff $\gamma = 25\%$.

\section{Benchmark Datasets}\label{data:bm}
We use the following benchmark datasets from \emph{Causeme} \citep{runge2019inferring,bussmann2021neural}.
\begin{enumerate}
    \item \underline{River Runoff.} This is a real dataset that consists of time series of river runoff at different stations. The time series have a daily time resolution and only include summer months (June-August). The physical time delay of interaction (as inferred from the river velocity) are roughly below one day, hence the dataset has contemporaneous time interactions. This dataset has 12 variables and 4600 time recordings for each variable.
    \item \underline{Logistic Map.} This is a synthetic dataset generated from logistic map with maximum time delay of 3 and a low dynamical noise and moderate strength of coupling between the variables. This dataset has 5 variables and 300 time recordings per variable.
\end{enumerate}

The PC algorithm is used to compute the DPGM from the scalar-valued random variables $Y_v$, defined by the average of recordings of a time window of length $\Delta t$, and averaging over alternate time windows of length $\Delta t$ and a gap of $\Delta t$ between consecutive windows results in samples of $Y_v$. Considering river runoff has large number of $4600$ time recordings while logistic map has only $300$ recordings, we used $\Delta t = 50$ and $3$ for river runoff and logistic map data respectively. The PC algorithm was implemented with p-value of $0.1$ for kernel-based non-linear conditional independence tests. In river-runoff and logistic map data, the TPC algorithm was implemented with maximum time-delay of interaction to be $1$ and $3$ recordings respectively, as per specification in the datasets, and a significance level $\alpha = 0.1$ for kernel-based conditional independence tests. For the bootstrap procedure in TPC, 50 bootstrap iterations with bootstrap window length of $50$ recordings and bootstrap stability threshold $\gamma = 0.01,0.15$ for river-runoff and logistic map datasets respectively.

\section{Visual Coding Neuropixels Dataset}\label{data:desc} For the purpose of application and comparison of the results of the methods discussed in this paper, we restrict our analysis to a 116 days old male mouse (Session ID 791319847) with 555 neurons whose spike trains are recorded simultaneously by six Neuropixel probes. The spike trains during the entire experiment were recorded at a frequency of $1$ KHz. We analyze the spike trains for four stimuli categories: 
\begin{enumerate}[leftmargin=*]
    \item Natural scenes, consisting of 118 natural scenes selected from three databases (Berkeley Segmentation Dataset, van Hateren Natural Image Dataset and McGill Calibrated Colour Image Database), with each scene presented briefly for 250ms and then replaced with the next scene image. Each scene is repeated 50 times in random order with intermittent blank intervals.
    \item Static gratings consisting of full-field sinusoidal gratings with 6 orientations (the angle of the grating), 5 spatial frequencies (the width of the grating), and 4 phases (the position of the grating) resulting in 120 stimulus conditions. Each grating is presented briefly (250 ms) before being replaced with a different orientation, spatial frequency and phase condition and each condition is repeated 50 times, in random order with intermittent blank intervals. 
    \item Gabor patches with 3 orientations where the patch center is lying at one of the points in a $9\times9$ visual field. Each gabor patch is being presented for 250ms and then replaced by a different patch, and each condition is repeated 50 times in random order with intermittent blank intervals.
    \item Full-field flashes, lasting for 250 ms followed by a blank interval of 1.75 s, and then the next flash, totaling 150 repetitions.
\end{enumerate}
This variety of stimuli is ranging from relatively \textit{natural stimuli} invoking mice's natural habitats (natural scenes) to \textit{artificial stimuli} (static gratings, gabors and flashes). Among the artificial stimuli, static gratings incorporate sinusoidal patches, while full-field flashes incorporate sharp changes in luminosity in the whole visual field in short period of time, and gabor patches incorporate sinusoidal patches with declining luminosity with distance from the center of the patch. With this choice of four stimuli we investigate how the variety of stimuli possibly invokes distinct patterns of neuronal interactions and connectivity. We exclude dynamic stimuli like natural movies, and drifting gratings, from this analysis because their results would require more nuanced study and interpretation, which we defer for future analysis.

\paragraph{Preprocessing}
We convert the spike trains recorded at 1 KHz to bin size of 10 ms by aggregating and then separating by start and end times of each stimuli presentation and obtain the Peri-Stimulus Time Histograms (PSTH) with bin-size 10 ms. We smooth the PSTHs by a Gaussian smoothing kernel of bandwidth $16$ms which provides a smoothed version of the PSTH for each neuron and each stimulus presentation. Some examples of the smoothed PSTH are displayed in Figure~\ref{fig:cfcneuropixels}. We use the smoothed PSTHs for neurons over each stimuli type as input for inference of the FC between the neurons for each stimuli presentation. For each stimulus presentation, we first selected the set of neurons that were active in at least $25\%$ of the bins in the PSTH, and then collected the set of unique neurons over all stimuli, which resulted in 54, 43, 19 and 36 active neurons for natural scenes, static gratings, gabor patches and flashes respectively, and 68 unique active neurons overall. We considered separated the entire duration of stimulus presentation to yield 58 trials of natural scenes, 60 trials of static gratings, 58 trials of gabor patches, and, 3 trials of flashes, where each trial is of duration 7.5 s.

We compare TPC with two popular methods for inferring the FC from neural signals: Granger Causality (GC) and Sparse Partial correlation via Graphical Lasso penalized Maximum Likelihood Estimation (Sparse Partial Corr). The TPC algorithm was implemented with maximum time-delay of interaction 10 ms, significance level $\alpha = 0.3$, $50$ bootstrap iterations, $250$ ms bootstrap window width, and stability threshold $\gamma = 0.01$. For GC, we consider VAR model of order $1$, and GC likelihood ratio statistic of greater than $90$ percentile as indicating edges \citep{schmidt2016multivariate}. For Sparse Partial Corr, the optimal penalization was determined by $5$-fold cross-validation. A summary of the results is provided in Figure~\ref{fig:cfcneuropixels}. 
\end{document}